\documentclass[12pt]{article}
\usepackage{amsmath,amsfonts, epsf,epsfig,color,latexsym,cite}
\usepackage{bbold}
\numberwithin{equation}{section}
\newcommand{\1}{\mathbb{1}}

\def\be{\begin{equation}}       \def\eq{\begin{equation}}
\def\ee{\label{abc}  \end{equation}}         \def\eqe{\label{abc}  \end{equation}}

\def\bea{\begin{eqnarray}}      \def\eqa{\begin{eqnarray}}
\def\ena{\end{eqnarray}}        \def\eea{\end{eqnarray}}
                                \def\eqae{\end{eqnarray}}

\def\a{\alpha}
\def\b{\beta}

\def\d{\delta}
\def\e{\epsilon}           
\def\f{\phi}               
\def\g{\gamma}
\def\h{\eta}
\def\i{\iota}

\def\l{\lambda}
\def\m{\mu}

  \def\w{\omega}
\def\p{\pi}                
\def\r{\rho}                                     
\def\s{\sigma}                                   

\def\x{\xi}

\def\F{\Phi}
\def\G{\Gamma}

\def\L{\Lambda}

\def\ca{{\cal A}}

\def\ce{{\cal E}}
\def\cf{{\cal F}}
\def\cg{{\cal G}}

\def\cl{{\cal L}}

\def\bop#1{\setbox0=\hbox{$#1M$}\mkern1.5mu
        \vbox{\hrule height0pt depth.04\ht0
        \hbox{\vrule width.04\ht0 height.9\ht0 \kern.9\ht0
        \vrule width.04\ht0}\hrule height.04\ht0}\mkern1.5mu}
\def\pa{\partial}                              
\def\we{\wedge}                                         
\def\>{\rangle} 

\def\<{\langle} 

\def\Tilde#1{\widetilde{#1}}                   

\def\to{\rightarrow}

\def\gij{g_{ij}}
\def\bij{b_{ij}}
\def\hijk{H_{ijk}}

\def\pa{\partial}

\def\ha{\frac12}                               

\def\IZ{\relax\ifmmode\mathchoice
{\hbox{\cmss Z\kern-.4em Z}}{\hbox{\cmss Z\kern-.4em Z}}
{\lower.9pt\hbox{\cmsss Z\kern-.4em Z}} {\lower1.2pt\hbox{\cmsss
Z\kern-.4em Z}}\else{\cmss Z\kern-.4em }\fi}
\def\IC{\relax\hbox{$\inbar\kern-.3em{\rm C}$}}
\def\IR{\relax{\rm I\kern-.18em R}}

\ifx\ltimes\undefined
\newcommand{\ltimes}{{\kern3pt\hbox{\vrule width 0.4pt height 5.30pt
depth .0pt}\kern-1.76pt\times\kern1pt}} \fi

\def\be{\begin{equation}}
\def\ee{\label{abc}  \end{equation}}
\def\ba{\begin{eqnarray}}
\def\ea{\end{eqnarray}}
\def\bq{\begin{quote}}
\def\eq{\end{quote}}

\def\part{\partial}

\def\beq{\begin{equation}}
\def\eeq{\label{abc}  \end{equation}}
\def\beqa{\begin{eqnarray}}
\def\eeqa{\end{eqnarray}}

\def\we{ \wedge}

\def\ti{\Tilde}

\def\Z {\mathbb{Z}}
\def\R {\mathbb{R}}

\def\XX {\mathbb{X}}

\def\cx{{\XX}}

\begin{document}
\thispagestyle{empty}
\begin{flushright}
hep-th/0604178\\
 Imperial/TP/06/CH/01 \\
 \end{flushright}\vskip 0.8cm\begin{center}
\LARGE{\bf   Global Aspects   of  T-Duality, Gauged Sigma Models  and T-Folds}
\end{center}
\vskip 1in

\begin{center}{\large C M  Hull }
\vskip 0.6cm{ Theoretical Physics Group,  Blackett  Laboratory, \\
Imperial College,\\ London SW7 2BZ, U.K.}\\
\vskip 0.8cm
and\\

\vspace*{7mm} {The Institute for Mathematical Sciences}\\
{\em Imperial College London} \\
{\em 48 Prince's Gardens, London SW7 2PE, U.K.} \\

\end{center}
\vskip 1.0cm

\begin{abstract}\noindent
The gauged sigma-model argument   that  string  backgrounds related by T-duality give equivalent quantum theories is revisited, taking careful account of global considerations.
The topological obstructions to gauging sigma-models give rise to obstructions to T-duality, but these are milder than those for gauging: it is possible to T-dualise a large class of  sigma-models that cannot be gauged.
For backgrounds that are torus fibrations, it is expected that T-duality can be applied fibrewise in the general case in which 
there are no globally-defined Killing vector fields, so that there is no isometry symmetry that can be gauged; the derivation of T-duality is extended to this case.
The T-duality transformations are presented in terms of globally-defined quantities. The generalisation to non-geometric string backgrounds is discussed,   the conditions for the T-dual background to be geometric found and the topology of T-folds analysed.

\end{abstract}

\vfill

\setcounter{footnote}{0}
\def\thefootnote{\arabic{footnote}}
\newpage

\section{Introduction}\label{Introduction}

The  two-dimensional sigma-model
 is a  theory of maps from a two-dimensional space $W$ to a manifold $M$ with a metric $g$ and closed 3-form $H$. Remarkably, in certain circumstances the two-dimensional quantum theory defined on
 $(M,g,H)$ can be the same as that defined by a sigma-model defined on a different manifold with different geometry   and topology $(\ti M, \ti g, \ti H)$. Of particular interest here  is T-duality, where 
  $(M,g,H)$ and the dual geometry $(\ti M, \ti g, \ti H)$ both have $d$ commuting Killing vectors with compact orbits 
  [1-19].
  The T-duality transformation from $M$ to $\ti M$ 
  can change the topology as well as the geometry \cite{Giveon:1993fd},\cite{Giveon:1993ph},\cite{Alvarez:1993qi},\cite{Bouwknegt:2003vb},\cite{Bouwknegt:2003wp}.  
  
If the target space of a   sigma model has isometries, the field theory has
 corresponding global symmetries. These can be promoted to   local symmetries  of the field theory by
 coupling to gauge fields on $W$, and such a gauged sigma model is the starting point for a 
 proof of the equivalence of the dual sigma models. In  
 \cite{buscher},\cite{Rocek:1991ps},\cite{Giveon:1991jj},  a gauged sigma model on a larger space is constructed
  with the extra coordinates appearing as  lagrange multipliers imposing the condition that the gauge fields are
pure gauge. Then two different gauge choices give  rise to two   sigma-models with different target spaces, but as they arise from two different ways of performing the same   path integral,
they give the same    quantum theory.

 However, it is not always  possible to gauge such an isometry symmetry:
  the  potential obstructions to gauging a sigma-model with non-trivial $H$  were found in \cite{Hull:1989jk},\cite{Hull:1990ms},\cite{Jack:1989ne}
and their topological interpretation explored in \cite{Hull:1990ms},\cite{Figueroa-O'Farrill:1994ns},\cite{Figueroa-O'Farrill:1994dj},\cite{Figueroa-O'Farrill:2005uz},\cite{Gualt}.
It is also not always possible to T-dualise  a sigma model with isometries, but the obstructions are weaker than those for gauging and there are ungaugable sigma-models that nonetheless can be T-dualised.
Many special cases have been discussed in the literature e.g.  \cite{Giveon:1991jj},\cite{Alvarez:1993qi},
\cite{Bouwknegt:2003zg},\cite{Bouwknegt:2004tr},
 but there does not seem to have been a general analysis.
The purpose here is to find the conditions necessary and sufficient conditions for a 
geometry   $(M,g,H)$ to have a geometric T-dual   $(\ti M, \ti g, \ti H)$, and also the conditions for there to be a T-dual with  a \lq non-geometric'  target space  \cite{Hull:2004in}. 
The conditons  found allow    a geometric T-dual to be found  for a  more general class of geometries than those discussed in  \cite{Giveon:1991jj},\cite{Alvarez:1993qi},
\cite{Bouwknegt:2003zg},\cite{Bouwknegt:2004tr}.
The local form of the transformations of course agree with those of  \cite{buscher},\cite{Rocek:1991ps},\cite{Giveon:1991jj}, and the 
novelty is in the understanding of global considerations.

An important example is that of a torus bundle in which there are local solutions to Killing's equations that generate the torus fibres, but which do not extend to globally defined vector fields.
In this case, there are no isometries, and so the analysis of  \cite{buscher},\cite{Rocek:1991ps},\cite{Giveon:1991jj} does not apply. Nevertheless, it is expected that one can apply duality fibrewise in such circumstances \cite{Vafa:1995gm}.
It will be shown here that there are potential obstructions to this, and when these are absent a gauged sigma-model derivation of the fibrewise T-duality will be given.
The discussion involves addressing the question of whether one can generalise the gauged sigma-model to the case of such torus bundles.

The action of the T-duality group $O(d,d;\Z)$ is usually presented in terms of fractional linear transformations of $g_{ij}+b_{ij}$, but there are problems with this if $b$ is only locally defined and is not a tensor field. One of the aims here will be to give a careful global characterisation of T-duality in terms of well-defined objects. This is an important pre-requisite to reformulating the results in terms of generalised geometry, as will be discussed elsewhere.

Suppose that $(M,g,H)$ has   $d$ (globally defined) commuting Killing vectors $k_m$, $m=1,...,d$, so that  $ \cl_m g=0$
where $\cl _m$ denotes the Lie derivative with respect to $k_m$, and
  that $H$ is invariant
\begin{equation} \label{LH}
{\cal L} _m H=0
\end{equation}
The Lie derivative of a form is given by
\begin{equation} \label{lie}
\cl _m  =\i _m d+ d \i _m
\end{equation}
where $\i _m$ is the interior product with $k_m$ (using the conventions of  \cite{Hull:1990ms})
so that (\ref{LH}) implies
\begin{equation} \label{abc}
d\i _m H=0
\end{equation}
and $\i_m H, \i_m \i_n H, \i_m \i_n \i_p H$ are closed forms on $H$.
The sigma-model action is invariant under  corresponding rigid symmetries provided
$\i _m H$ is exact, so that
\begin{equation} \label{ihis}
\i _m H=dv_m
\end{equation}
for some globally-defined 1-forms $v_m$ \cite{Hull:1989jk}.

Given a suitable good open cover $\{U_{\alpha}\}$ of the manifold $M$  (in which each $\{U_{\alpha}\}$ has trivial cohomology), in 
  each patch $U_\a$ a two-form $b^\a $ can be found such that
\begin{equation} \label{abc}
H=db^\a
\end{equation}
In the overlap $U_{\alpha}\cap U_{\beta}$, the difference between the $b$'s must be closed and so exact, so that
\begin{equation}\label{abc}
b^\a - b^\b = d \L ^{\a\b}
\end{equation}
for some one-form $ \L ^{\a\b}$ in $U_{\alpha}\cap U_{\beta}$ (satisfying the usual consistency condition in triple overlaps).
Then $b^\a$ is a local potential for the field strength $H$, 
and is determined by $H$ up to local gauge transformations
\begin{equation} \label{abc}
\d b^\a = d \l   ^\a
\end{equation}
where  $\l ^\a $ is a one-form on $U_\a$.
The potential $b$ need only be invariant up to a gauge transformation, so that
\begin{equation} \label{abc}
\cl_m b ^\a=d w_m ^\a
\end{equation}
for a 1-form $w_m ^\a$ in $U_\a$ given by
\begin{equation}  \label{abc}
w_m ^\a= v_m +\i_m b^\a
\end{equation}

To be able to T-dualise using the $d$ Killing vectors requires that the
orbits be compact, so that $M$ has a torus fibration with fibres $T^d$.
In \cite{Giveon:1991jj}, T-duality was analysed for the case in which a gauge can be chosen in which
$\cl _m b^\a=0$. However, such a gauge  is not possible for all patches in  general. For example, such a gauge choice cannot be possible  if there is non-trivial $H$-flux on the fibres (i.e. if $\int H$ is non-zero over a cycle of the $T^d$ fibres). 
In \cite{Alvarez:1993qi}, a global derivation of T-duality was given  for one Killing vector  ($d=1$) for the case in which $\i _m H$ is exact.
   It was then argued that  this condition can be relaxed by choosing  coordinate patches on $M$ in which 
    $\i _m H$ is exact in each patch, and then patching together the gaugings from the different patches.
    This was shown to work in some interesting examples, but the questions as to whether such a patching is always possible and whether this extends to more than one Killing vector were not addressed.
    
   In \cite{Bouwknegt:2003zg},\cite{Bouwknegt:2004tr}
 the case of principle torus bundles was discussed. Dimensional reduction of $H$ on the $T^d$ fibres gives forms $H_3,H_2,H_1,H_0$ where $H_p$ is a  $p$-form  on the base. 
It was claimed that T-duality was possible if $H_1=0, H_0=0$ and that otherwise there is an obstruction.
  There is also a 2-form $F_2$ on the base which is the curvature of the connection on the bundle, and both $H_2$ and $F_2$ take values in the Lie algebra of $U(1)^d$.
  The topology is characterised by two
  integral cohomology classes on the base, the first Chern class $[F_2]$ and the \lq $H$-class'  $[H_2]$, and T-duality interchanges the two, so that  $[\ti F_2]=[H_2]$ and 
  $[\ti H_2]=[F_2]$.

Here the general case of simultaneous T-duality in $d$   directions will be analysed, for general $T^d$ fibrations  (i.e. $M$ need not be a   principle torus bundle). 
In this article, only the case in which the local $U(1)^d$ acts without fixed points will be discussed.
First, in the case of $d$ globally-defined nowhere-vanishing Killing vector fields, the result is  that the condition for a geometric T-duality to be possible, i.e. one in which   the dual is   again a manifold $\ti M$
with tensor fields $\ti g, \ti H$,
are that the closed 2-form $\i _m H$ is the curvature for some line bundle, that
$\i _m \i_n H$ is exact
and that  $\i _m \i_n \i_p H=0$.
 This includes cases in which 
$\i _m  H$ is not exact, so that the original sigma-model is not invariant under the action of $U(1)^d$, and in which $H_1$ is non-zero. 
This is then generalised to the case of torus bundles, 
where a modification of the constraint on $\i _m \i_n H$ is found, while
 $\i _m \i_n \i_p H=0$ is still needed.
The general form of the T-duality transformations are given 
in terms of globally-defined geometric structures -- of course, they agree with those given  in \cite{buscher},\cite{Giveon:1991jj}
 locally.

An important question is whether T-duality is possible under   more general circumstances.
In \cite{Hull:2004in} it was argued that in certain cases the T-dual is a T-fold -- a space which looks locally like a manifold with $g,H$
but 
where the transition functions between patches involve T-duality transformations.
Examples of such non-geometric string backgrounds have been explored in \cite{Hull:2004in},[27-34].
It will be shown here that the only condition for a T-duality to a T-fold to be possible is that
the constants $\i _m \i_n\i_p H$ vanish, and no condition on $\i _m \i_n H$ is  needed.
In \cite{Dabholkar:2005ve}, it was argued that T-duality of more general cases with $\i _m \i_n\i_p H\ne 0$
is in fact possible, with a result that is a stringy geometry that does not look like a conventional manifold even locally. (An alternative viewpoint was taken in  
 \cite{Mathai:2004qq},\cite{Mathai:2005fd}, \cite{Bouwknegt:2004ap}.
It was argued that if $H_1\ne 0$ and $H_0=0$ the dual is a non-commutative geometry in \cite{Mathai:2004qq},\cite{Mathai:2005fd}
and that if $H_0\ne 0$ then it is a non-associative geometry  in \cite{Bouwknegt:2004ap}.)
 
The plan of the paper is as follows. In section 2, a review is given of the gauging of sigma-models with Wess-Zumino term and in particular of the obstructions to gauging. Section 3 further examines the geometry of 
manifolds  $M$  that are torus bundles, with a metric $g$ and closed 3-form $H$ that are invariant under a $U(1)^d$ group action,
and in particular investigates the quotient  geometry arising from the integrating out the gauge fields in the corresponding gauged sigma-model with WZ term. There are problems with the usual formulation of the T-duality transformations; for example, they involve non-linear transformations of the 2-form gauge field which appear inconsistent with the 2-form gauge symmetry. Geometric quantities are introduced in terms of which T-duality can be expressed covariantly.

Section 4 shows that almost all of the obstructions to gauging a $U(1)^d$ group action
can be overcome by introducing a further $d$ scalar fields. Geometrically, these extra scalars correspond to the fibre coordinates of a $d$-torus bundle $\hat M$  over $M$, which is the  doubled torus of \cite{Hull:2004in}. These extra scalars can also be thought of as the extra lagrange multiplier fields introduced in the sigma-model derivation of T-duality \cite{buscher},\cite{Rocek:1991ps},\cite{Giveon:1991jj}. In section 5, the global structure of $\hat M$ is analysed, and in particular the periodicities of the extra coordinates shown to be inversely related to the periodicites of the 
fibre coordinates of $M$. It is seen that there are some subtleties in identifying precisely which are the correct periodic coordinates.

Section 6 uses the results from the previous sections to re-examine the sigma-model derivation of T-duality. The standard  derivation   gauges an abelian isometry and  adds lagrange multiplier fields constraining the gauge fields to be trivial. Then integrating out the lagrange multipliers  and gauge-fixing recovers the original geometry while integrating out the gauge fields gives the T-dual geometry. Section 6 generalises this to a wide class of geometries where the first step of gauging the sigma-model is not possible, and in this way it is seen that the obstructions to T-duality are considerably weaker than the obstructions to gauging a sigma-model. Nevertheless, there are some obstructions to T-duality and these are carefully discussed. The T-duality transformations are expressed covariantly in terms of geometric variables.

Section 7 examines more general torus bundles in which there is no action of $U(1)^d$, These are not principle bundles, and although Killing vectors exist locally, they do not extend to global vector fields.
The adiabatic argument suggests that  T-duality can be applied fibrewise in such situations, even though the general T-duality derivation of section 6 fails in this case. A more general construction is proposed that formally establishes fibrewise T-duality in this case.
Section 8 looks at a more general set-up in which the transition functions involve B-shifts. Local application of the T-duality rules lead to a set of patches of dual geometry that cannot fit together into a geometric background but which do fit together to form a non-geometric background, a T-fold. However, a derivation of this result using gauged sigma-models is not possible.
In section 9, the  discussion  of T-duality is  extended to T-folds.

\section{Gauged Sigma Models}\label{Gauged}

The sigma model with target space $M$ is a theory of maps $\f:W\to M$. If $X^i$ are coordinates on $M$ and $\s ^a$ are coordinates on $W$, the map is given locally by functions $X^i(\s)$.
The action is the sum of a kinetic term $S_g$ and a Wess-Zumino term $S_{WZ}$
\begin{equation}
S_0=S_g+S_{WZ}
\end{equation}
Given a metric $g$ on $M$, the kinetic term is 
\begin{equation} \label{abc}
S_g=\ha  \int _W   \gij \,   dX^i  \wedge *dX^j
\end{equation}
Here and in what follows, the pull-back $\f ^*(dX^i )= \pa _a X^i d \s ^a$ will be written $dX^i$,
and it should be clear from the context whether a form on $M$ or its pull-back is intended.
 The Hodge dual on $W$ constructed using a metric $h_{ab}$   is denoted $*$.

The Wess-Zumino term
is constructed using a closed 3-form $H$ on $M$.
If $H$ is exact, then
  there is a globally defined 2-form $b$ on $M$
  with
  \begin{equation}
H=db
\end{equation}
  and   the Wess-Zumino term is
\begin{equation} \label{wzb}
S_{WZ}=     \int  _W  \f ^*b = \ha  \int _W   \bij  \,  dX^i  \wedge dX^j
\end{equation}
This can be rewritten as
\begin{equation} \label{wzh}
S_{WZ}=  \int  _V  \f ^*H= \frac13 \int _V \hijk\,  dX^i  \wedge dX^j\we dX^k
\end{equation}
where $V$ is any 3-manifold with boundary $W$.

This form of the action can also be used in
the case in which $H$ is not exact. Then the action
depends on the choice of $V$, but the difference between the actions for two choices   $V,V'$ with the same boundary $W$ is
\begin{equation}
S_{WZ}(V)-S_{WZ}(V')= \int  _{V-V'}  \f ^*H = \int _{\f(V-V')}H
\end{equation}
where $V-V'$ is the compact 3-manifold obtained from glueing $V$ to $V'$ along their common boundary
with opposite orientations, and ${\f(V-V')}$ is the corresponding closed 3-manifold in $M$.
The result is a topological number depending only on the cohomology class of
$H$ and the homology class of   ${\f(V-V')}$, so that the choice of $V$ does not affect the classical field equations.
The ambiguity in the choice of $V$ leads to an ambiguity in the Euclidean functional integral $\int [dX] \exp {(-k S)}$
by a phase
\begin{equation}
\exp {ik  \int _{\f(V-V')} H}
\end{equation}
where $k$ is a coupling constant.
The functional integral is then well-defined provided
$\frac {k}{2\p} [H]$ is an integral cohomology class (where $[H]$ is the de Rham cohomology class represented by $H$).

Suppose there are $d$ commuting Killing vectors $k_m$ with $\cl_m H=0$.
Then under the transformation
\begin{equation} \label{trans}
\d X^i = \a^m k_m^i(X)
\end{equation}
with constant parameter $\a^m$ 
the action changes by
\begin{equation} \label{abc}
\d S=     \int  _W  \f ^*(\a^m \i _m H)
\end{equation}
and this will be a surface term if
$\i _m H$ is exact, so that
\begin{equation} \label{vis}
\i _m H=dv_m
\end{equation}
for some (globally defined) one-forms $v_m$, which are defined by (\ref{vis}) up to the addition of   exact forms \cite{Hull:1989jk}. This is then a global symmetry if $\i _m H$ is exact.

Gauging of the sigma-model \cite{Hull:1989jk},\cite{Jack:1989ne} consists of promoting the symmetry (\ref{trans}) to a local one with   parameters that are functions  $\a^m(\s)$
by seeking a suitable coupling to   connection one-forms $C^m$ on $W$
transforming as
\begin{equation} \label{ctrans}
\d C^m =d\a^m
\end{equation}
It was shown in  \cite{Hull:1989jk},\cite{Jack:1989ne}  that gauging is possible if  
 $\i_mH$ is exact, and    a one-form $v_m=v_{mi} dX^i$ can be chosen with $\i _m H=dv_m$ that satisfies
\begin{equation} \label{liev}
\cl _m v_n=0
\end{equation}
(so that  $\i _m H$ represents a  trivial equivariant cohomology class)
 and
 \begin{equation} \label{ivas}
\i_m v_n =-\i_n v_m
\end{equation}
This defines globally-defined functions
 \begin{equation} \label{bis}
B_{mn}=\i_m v_n  
 \end{equation}
 satisfying $B_{mn}=-B_{nm}$ and $\cl_p B_{mn}=0$.
 The identity
 \begin{equation} \label{iihislv}
\i_m \i_n H= \cl _ mv _n - d \i _m v_n
\end{equation}
together with $\cl _m v_n=0$
implies  $\i_m \i_n H$ is exact with
\begin{equation} \label{iihis}
\i_m \i_n H= -d B_{mn}
\end{equation}
Finally
\begin{equation} \label{iiihis}
\i_m \i_n \i_p H= 0
\end{equation}
as  $\cl_p B_{mn}=0$.

The covariant derivative of $X^i$ is
\begin{equation} \label{dcis}
D_a X^i =\pa _a  X^i - C_a ^m k_m^i
\end{equation}
with field strength
\begin{equation} \label{abc}
\cg ^m=dC^m
\end{equation}
The gauged action is
\cite{Hull:1989jk}
\begin{equation} \label{gag3}
S= \frac12 \int  _W    \gij  D  X^i \we *D  X^j +   \int  _V   \left(  \frac13 \hijk   D  X^i \we D  X^j \we D X^k  +  \cg ^m  \we  v_{mi} DX^i \right)
\end{equation}
which can be rewritten as (choosing a flat metric $h_{ab}=\h_{ab}$) \cite{Hull:1989jk},\cite{Jack:1989ne} 
\begin{equation} \label{gag2}
S_0 + \int _W \left(- C^m_a J^a_m + \ha C^m_aC^n_b\left[ G_{mn} \h^{ab}  + B_{mn} \e ^{ab}\right]
\right)
\end{equation}
where $S_0$ is the ungauged action,
\begin{equation}
G_{mn}=g_{ij}k^i_mk^j_n
\end{equation}
 and
\begin{equation} \label{jis}
J^a_m=( k_{mi} \h^{ab}- v_{mi} \e^{ab}) \pa _b X^i 
\end{equation}
Introducing light-cone world-sheet coordinates $\s^a =( \s^+, \s^-)$ with $ \h^{+-} = \e ^{+-}=1$, this can be rewritten as
\begin{equation} \label{gagc}
S_0 + \int _W \left( -C^m_+ J^+_m - C^m_- J^-_m + C^m_+E_{mn}C^n_- \right)
\end{equation}
where
\begin{equation} \label{eiss}
E_{mn}=G_{mn} + B_{mn}
\end{equation}
and
\begin{equation} \label{jlc}
J_{m\pm} = (k_{mi} \pm v_{mi}) \pa _\pm X^i
\end{equation}
The ungauged action can be written as
\begin{equation} \label{epact}
\int _W d^2\s \, \, \ce _{ij} \pa _+ X^i \pa _- X^j
\end{equation}
where
\begin{equation}
\ce _{ij}= \gij + \bij
\end{equation}

If $E_{mn}(X)$ is invertible for all $X$, then writing 
the gauge fields $C=\ti C + \F$ where
\begin{equation} \label{ctiis}
 \ti C_+= (E^t)^{-1} J_+, \qquad \ti C_-=E^{-1} J_-
\end{equation}
 gives
\begin{equation} \label{sdu}
S'=S_0- \int _W d^2\s \,  J^-_m (E^{-1})^{mn} J^+_n
\end{equation}
plus 
\begin{equation} \label{fact}
S_\F=  \int _W d^2\s \,  \F^m_+E_{mn}\F ^n_-
\end{equation}
Note that $\ti C $ transforms as a gauge field under the local transformations (\ref{trans}) $\d \ti C= d \a$ \cite{Hull:1989jk}, so that 
$\F _a ^m $ are globally-defined world-sheet vectors.
The action $S_\F$ involves no derivatives so that the $\F$ are auxiliary fields with no dynamics.
The action (\ref{sdu}) can be written as
\begin{equation} \label{epact}
\int _W d^2\s \, \,  \ce '_{ij} \pa _+ X^i \pa _- X^j
\end{equation}
where
$\ce _{ij}$ has been transformed to
\begin{equation}\label{epis}
\ce '_{ij}=\ce _{ij}- (k_{mi}+v_{mi} )(E^{-1})^{mn}  (k_{mj}-v_{mj} )
\end{equation}
This amounts to gauging using the connection
$\ti C$, and so is automatically invariant under the local transformations (\ref{trans}).

If  the isometry acts without fixed points and if  $g_{ij}$ induces a positive-definite metric on the fibres, then $G_{mn}$ is invertible.
The matrix  $E$ is degenerate at points $X_0$ at which there is a   vector  $U$ such that
$E(X_0)U=0$, so that $G_{mn}U^n=-B_{mn}U^n$. This implies that
$G_{mn}U^mU^n=0$ so that at $X_0$ there is a Killing vector $K$ (some linear combination of the $k_m$) that becomes  null. For positive definite $G_{mn}$, this implies   $K(X_0)=0$ so that $X_0$ is a fixed point for $K$. Then $E$ is invertible if and only if the isometry group acts without fixed points.

\section{The Geometry of Gauged Sigma Models}

Suppose the  abelian isometry group $G$ generated by the Killing vectors   acts without fixed points. 
Then the quotient    $M/G$
 defines the space of orbits $N$, and is a manifold. As a result,  $M$ is a bundle over $N$ with fibres $G$, with projection $\pi:M\to N$.
A form $\w$ satisfying $\i _m \w=0$ will be said to be {\it horizontal}, one satisfying 
$\cl_m \w=0$ will be said to be {\it invariant} and one that is both horizontal and invariant is 
{\it basic}.  Equivariant cohomology is the cohomology of basic forms, and
the obstructions to gauging can be characterised  in terms of this cohomology \cite{Figueroa-O'Farrill:1994ns},
\cite{Figueroa-O'Farrill:1994dj},
\cite{Figueroa-O'Farrill:2005uz}.
A metric $g$ on $M$ will be said to be horizontal if the Killing vectors $k_m$ are null and satisfy
$g(k_m, V)=0$ for all $V$, and a horizontal metric which is  invariant $(\cl g=0)$ will be said to be basic.
Basic metrics and forms on $M$ can be thought of as metrics and forms on $N$, as  they are the images
under the pull-back $\pi ^*$ of metrics and forms on $N$.

\subsection{A Single Killing Vector}

Before proceeding to the general case, it will be useful to discuss the case $d=1$  with one Killing vector $k$.
Let  $G= \gij k^i k^j$, and it will be assumed that $G$ is nowhere vanishing (so that
there are no fixed points).  Then  $M$ is a line or circle bundle over   some manifold $N$, with 
fibres given by the orbits of $k$.
  It is useful to define the dual one-form $\x$ with components $\x _i = G^{-1}\gij k^j $, so that
  $\i \x =1$ where $\i $ is the interior product with $k$.
  The 2-form
  \begin{equation}
F=d\x
\end{equation}
is horizontal
\begin{equation}
\i F=0
\end{equation}

The metric takes the form
\begin{equation} \label{ghat}
g= \bar g + G \, \x \otimes \x
\end{equation}
where $ \bar g(k, \cdot)=0$
so that $\bar g$  is basic and can be thought of as  a metric on the quotient space $N$.
In adapted local coordinates 
$X^i= (X, Y^\m)$ in which 
\begin{equation} \label{abc}
k^i \frac {\pa} {\pa X^i}= \frac {\pa} {\pa X}
\end{equation}
and $Y^\m$ are coordinates on $N$, 
the Lie derivative is the partial derivative with respect to $X$, so that
$\gij,\hijk$ are independent of $X$.
Then 
\begin{equation} \label{abc}
\x = dX + A
\end{equation}
where
$A=A_\m (Y)d Y^\m$ satisfies
$\i  A=0 $ and
\begin{equation} \label{abc}
dA=F
\end{equation}
Then $A $ is a connection 1-form for $M$ viewed as a bundle over $N$.

If the symmetry is gaugable, there is a globally defined $v$ with $\i H=dv $
and \begin{equation}
\i v=0, \qquad \cl _k v=0
\end{equation}
Then
\begin{equation}
\ti F = dv
\end{equation}
is also horizontal, $\i \ti F =0$. In the adapted coordinates,
$v=v_{\m }d Y^\m$.

The 3-form $H$ can be decomposed as 
\begin{equation} \label{abc}
H=h +(\i H)  \we dX =h +(dv)  \we dX
\end{equation}
where $h$ is a horizontal closed 3-form, 
$\i h=0$ and  $dh=0$.
As a result $H=db$ where
\begin{equation}
b= \bar b +v\we dX
\end{equation}
and  $h=d\bar b$.
There are similar expressions using $\x$ instead of $dX$ 
\begin{equation}
H= \bar H+ dv\we \x= \bar H +  \ti F\we  \x
\end{equation}
where
\begin{equation}
\bar H= d\bar b -  \ti F\we A
\end{equation}
satisfies
\begin{equation}
d\bar H= - F\we \ti F
\end{equation}
and is horizontal, $\i \bar H=0$, and so basic.
Here $\bar H$ is a globally defined 3-form.

If the orbit of $M$ is a circle so that $M$ is a circle bundle, the topology is characterised by 
the first Chern class, $[F]\in H^2(N)$.
The topology associated with the $b$-field is characterised by the cohomology class $[\ti F]\in H^2(N)$, and this will be referred to as the $H$-class. It will be seen in section 5 that, when appropriately normalised, both correspond to  integral cohomology classes.

Next, consider the geometry $(M,g',H')$ obtained by gauging $k$ and eliminating the gauge field. It is given by 
(\ref{epis}), which implies
\begin{equation}
\ce ' _{ij}= \ce_{ij} - (G\x _i +\bar  \x_j)G^{-1} (G\x _j-\bar  \x _j)
\end{equation}
and the notation
$\bar \x_i \equiv v_i$ has been introduced for comparison with later formulae.
The symmetric and anti-symmetric parts give
\begin{equation}
g'_{ij}= \gij - G \x_i\x_j + G^{-1}\bar  \x_i\bar  \x_j, \qquad b'_{ij}= \bij -\bar  \x_i\x_j +\x_i\bar  \x_j 
\end{equation}
Then
\begin{equation}
\label{fdhkasf}
  g' = g - G \, \x \otimes \x +G^{-1} \, \bar  \x \otimes \bar  \x =\bar g  +G^{-1} \, \bar  \x \otimes \bar  \x
\end{equation}
and 
 \begin{equation}
 \label{kjdflkadjflk}
  H' =  H- \ti F \we \x +\bar  \x  \we F =\bar H+ \bar  \x  \we F
\end{equation}
 are both horizontal, using $\i \bar  \x=0$,
 \begin{equation}
\i H'=0, \qquad g' (k, \cdot )=0
\end{equation}
 as well as invariant.
  This is sufficient to ensure that
 $\d X^i = \a k^i$ is a symmetry of the sigma model on
 $(M,g',H')$. The Killing direction is null for the metric $g'$.
 One can then take the quotient with respect to the isometry to obtain a sigma model on the quotient space $N$, with geometry $(N,g',H')$.
 More physically, the local  symmetry can be fixed by choosing $X(\s ) = X_0$
 for some point on the orbit   and the sigma model reduces to one on $N$ with coordinates
 $Y^\m$. This amounts to choosing a section of the bundle, and in general there will not be a global section, so that one may need to choose   different gauge choices $ X_0$ over each patch in $N$.

\subsection{Several Killing vectors}

Consider $(M,g,H)$ with $d$ commuting Killing vectors, and 
suppose that   $G_{mn}$ and $E_{mn}$ are invertible everywhere.
It is useful to define the one-forms 
$\x^m$ with components
\begin{equation} \label{xis}
\x^m _i=G^{mn} g_{ij} k_n^j
\end{equation}
so that they are dual to the Killing vectors
\begin{equation} \label{abc}
\x^m(k_n)=\d ^m{}_n
\end{equation}
and
satisfy
\begin{equation} \label{abc}
\i_m  F^n=0
\end{equation}
where
\begin{equation} \label{abc}
F^m=d\x^m
\end{equation}
The metric can be written as
\begin{equation} \label{gis}
g = \bar g +   G_{mn}\, \x^m \otimes \x^n
\end{equation}
where  $\bar g$ is 
a  basic metric
with
$\bar g(k_m, \cdot)=0$ 
so that it can be viewed as a  metric on $N$.

In adapted local coordinates 
$X^i= (X^m, Y^\m)$ in which 
\begin{equation} \label{abc}
k_m^i \frac {\pa} {\pa X^i}= \frac {\pa} {\pa X^m}
\end{equation}
the Lie derivative is the partial derivative with respect to $X^m$, so that
$\gij,\hijk$ are independent of $X^m$.
Then 
\begin{equation} \label{aiss}
\x^m = dX^m + A^m
\end{equation}
where
$A^m=A^m_\m (Y)d Y^\m$ satisfies
$\i _m A^n=0 $ and
\begin{equation} \label{abc}
dA^m=F^m
\end{equation}
satisfies 
$\i_m F^n=0$.
The $A^m $ are connection 1-forms for $M$ viewed as a bundle over $N$.

Any form on $M$ can be expanded using either the forms $dX^m$ defined in a local coordinate patch, or using the globally-defined one-forms $\x^m$.
From (\ref{bis}), 
\begin{equation} \label{vbis}
v_m= -B_{mn} \x^n + \bar  \x_m
\end{equation}
for some globally-defined basic  one-form $\bar  \x_m$.
Defining the basic 2-form
\begin{equation}\label{ftiiss}
\ti F_m = d \bar  \x _m
\end{equation}
one has
\begin{equation}
dv_m = \ti F_m-B_{mn} F^n - dB_{mn} \we \x^n
\end{equation}
Note that $dB_{mn}$ is basic.
The 1-forms $\bar \x$ are given in terms of $v$ by
\begin{equation}
\label{}
\bar \x _m = [v_m - (\i _n v_m) \x ^n]  +(B_{mn} + \i _n v _m) A^n
\end{equation}

 The 3-form $H$ can be written as
\begin{equation} \label{hiss}
H=\bar H +(\i_m H) \we  \x^m +\ha ( \i _m \i _n H) \we \x ^m  \we \x ^n - \frac 16 (  \i _m \i _n \i _pH)  \we \x ^m  \we \x ^n \we  \x ^p
\end{equation}
where
$\i _m \bar H=0$.
Using (\ref{vis}),(\ref{iihis}),(\ref{iiihis}) this becomes
\begin{equation} \label{abc}
H=\bar H+ (dv_m) \we   \x^m -\ha (d B_{mn}) \we  \x ^m \we  \x ^n 
\end{equation}
giving
\begin{equation} \label{abc}
H=\bar H + ( \ti F_m-B_{mn} F^n ) \we   \x^m + \ha (d B_{mn}) \we  \x ^m \we  \x ^n 
\end{equation}
or equivalently
\begin{equation} \label{hiso}
H=\bar H +  \ti F_m \we   \x^m + dB
\end{equation}
where
\begin{equation}\label{biso}
B=\ha   B_{mn}
   \x ^m \we  \x ^n 
\end{equation}
is a globally-defined 2-form.
Closure of $H$ requires that $\bar H$ satisfy
 \begin{equation}
d\bar H= - \ti F_m \we F^m
\end{equation}

As $\bar H$ is basic and $\bar H +  \ti F_m \we   \x^m $ is closed, 
\begin{equation}
\label{}
\bar H +   F^m \we   \bar \x_m = \bar H +  \ti F_m \we   \x^m + d(\x^m \we \bar  \x _m)
\end{equation}
is closed and basic, and so locally this is $d \bar b $ where
 $\bar b $  is a basic 2-form. Then locally $H=db$ where
 \begin{equation}
\label{bisoo}
b= \bar b +  \x^m \we \bar  \x _m + B
\end{equation}
and
\begin{equation}
\label{rtrhrthrmh}
\bar H = d \bar b -    F^m \we   \bar \x_m 
\end{equation}
There are now $d$ 1st Chern classes $[F^m]
\in H^2(N)$
and $d$ {}$H$-classes $[ \ti F_m]\in H^2(N)$.

 Consider now the geometry   $(M,g',H')$ arising from  eliminating  $C$, given by (\ref{epis}).
 Rewriting in terms of $\x,\bar  \x$, a remarkable simplification occurs.
 The equations  (\ref{xis}),(\ref{vbis}) imply
 \begin{equation}
k_{mi}-v_{mi}= E_{mn} \x^n -\bar  \x _m, \qquad k_{mi}+ v_{mi}= E_{nm} \x^n + \bar  \x _m
\end{equation}
so that
\begin{equation}
\label{ }
J_-= (E_{mn} \x^n_i - \bar \x _{mi})\pa _-X^i, \qquad J_+= (E_{nm} \x^n_i + \bar \x _{mi})\pa _+X^i
\end{equation}
and the induced connections $\ti C$ are
\begin{equation}
\label{ }
\ti C_-^m= ( \x^m_i -(E^{-1})^{mn} \bar \x _{ni})\pa _-X^i, \qquad \ti C_+^m= ( \x^m_i +(E^{-1})^{nm} \bar \x _{ni})\pa _+ X^i\end{equation}
Using (\ref{aiss}), this can be rewritten as
\begin{equation}
\label{ }
\ti C^m _a= A^m_i \pa _a X^i + \F^m_a
\end{equation}
where $\F^m_a$ is a globally-defined one form on $W$ constructed 
using $\bar \x$, plus a pure gauge term $\pa_aX^m$.
Thus the connections $C$ and $\ti C$ on $W$ 
are given by 
  the pull-back of the connection $A$ on the bundle $M\to N$, plus   global one-forms, so that the $U(1)^d$ bundle over the world-sheet is the pull-back of the torus bundle over $N$.

The new geometry obtained by integrating out the gauge fields is  given by
\begin{eqnarray}
\ce '_{ij}&=&\ce _{ij}- (E_{pm}\x_{i}^p+ \bar  \x_{mi} )(E^{-1})^{mn}  (E_{nq}\xi_{j}^q-\bar  \x _{nj} )
\nonumber \\
 & = &  \ce _{ij}- \x_{i}^mE_{mn}\xi_{j}^n +\bar  \x_{mi} (E^{-1})^{mn}\bar  \x _{nj}
 -\bar  \x_{mi}\xi_{j}^m+\x_{i}^m\bar  \x _{mj}
\end{eqnarray}
Defining the symmetric and anti-symmetric parts  
\begin{equation}
\ti G^{mn}= (E^{-1})^{(mn)}, \qquad \ti B^{mn}= (E^{-1})^{[mn]}
\end{equation}
the geometry is given by
\begin{eqnarray}
g' & = & g - G_{mn} \x ^m \otimes \x ^n + \ti G^{mn} \bar  \x _m\otimes  \bar  \x _n \\
b' & = & b - \bar  \x_m \we \x ^m  - \x_{i}^mB_{mn}\xi_{j}^n +\bar  \x_{mi} \ti B^{mn}\bar  \x _{nj}
\end{eqnarray}
Using (\ref{gis}),
\begin{equation}
g'=\bar g  + \ti G^{mn} \bar  \x _m\otimes \bar  \x _n
\end{equation}
while
\begin{equation}
H' = H  - \ti F _m\we \x^m + \bar  \x _m \we  F^m
\end{equation}
so that from (\ref{hiso}),(\ref{biso})
\begin{equation} \label{abc}
H'=\bar H + \bar  \x _m \we  F^m + d\ti B
\end{equation}
where
\begin{equation}
\ti B=\ha \ti   B^{mn}
   \bar  \x _m \we  \bar  \x _n 
\end{equation}
Thus the gauging together with elimination of gauge fields leads to the changes
   \begin{eqnarray}\label{kdjjsfh}
g & = & \bar g + G_{mn} \x ^m \otimes \x ^n \to 
g'=\bar g  + \ti G^{mn} \bar  \x _m\otimes \bar  \x _n \\
\label{kdjjsfh2}
H&=&\bar H +  \ti F_m \we   \x^m + dB
\to H'=\bar H + \bar  \x _m \we  F^m + d\ti B
\end{eqnarray}
  which then interchanges $\x$ with $\bar  \x$ and takes $E\to E^{-1}$. 
    
 Note that $g',H'$ are invariant and  horizontal with respect to all of the Killing vectors
 \begin{equation}
\i _m H'=0, \qquad g' (k_m ,\cdot)=0
\end{equation}
so that the sigma-model on $(M,g',H')$ is invariant under the local symmetries
$\d X^i = \a ^m k_m^i$.
 This can be checked directly, or by noting that
 eliminating any one of the $C_m$ gives a geometry that is horizontal with respect to the corresponding Killing vector, and then repeating the argument for each of the $d$ gauge fields in turn.
Again one can take the quotient under the action of the isometry group to obtain a sigma model on $(N,g',H')$.
This can be thought of as fixing the symmetry by choosing   local sections of the bundle,
   fixing all of the coordinates $X^m$, 
 so that  the sigma model reduces to one on $N$ with coordinates
 $Y^\m$.
 
 \subsection{Global Symmetries}
 
 Suppose the orbits of each of the $k_m$ are periodic, so that $M$ is a torus bundle over $N$.
 The general Killing vector with periodic orbits is of the form  $\sum_m N^m k_m$ 
 where $N^m$ are integers.
 One can then change from the basis $\{k_m\}$ to a new basis $\{k'_m\}$ of Killing vectors with periodic orbits
 \begin{equation}
k'_m = L_m {}^n k_n
\end{equation}
where $ L_m {}^n$ is any   matrix in $GL(d,\Z)$.
The components of $G_{mn}, B_{mn}, \x^m, v_m$ in the new basis are then
\begin{equation}
G'=  LGL^t, \qquad B'=  LBL^t, \qquad \x' = (L^t)^{-1}\x, \qquad v'=Lv
\end{equation}
This gives a natural action of $GL(d,\Z)$ in which     upper  indices $m$ transform in the  vector representation and lower indices transform in the co-vector representation.
The periodic coordinates $X^m$ adapted to $k_m$ and the coordinates ${X'}^m$ adapted to $k'_m$
with
\begin{equation}
k_m=\frac {\pa} {\pa X^m}, \qquad
k'_m=\frac {\pa} {\pa {X'}^m}
\end{equation}
are related by 
\begin{equation}
{X'}^m= (L^{-1})_n{}^m X^n
\end{equation}
which is a large diffeomorphism of the torus.

The metric and $b$-field are given in terms of $G_{mn}, B_{mn}, \x^m, v_m$. Then 
$G',B',v',\x'$ determine the same geometry as $G,B,v,\x$ if they are related by a $GL(d,\Z)$ transformation, as one is transformed to the other by a change of basis. 
Then $GL(d,\Z)$  is a symmetry, as target spaces related 
by the action of  $GL(d,\Z)$  are equivalent and
determine the same physical models.

A shift  
\begin{equation}
B_{mn} \to B_{mn} + \b_{mn}
\end{equation}
where $\b_{mn}$ are constants leaves $H$ unchanged and so the classical physics is unaltered.
The action changes by
\begin{equation}
\ha \int _W \b_{mn} dX^m \we dX^ n = \int _{\f (W)} \b
\end{equation}
which is the integral of the 2-form $\b$ over the embedding of the world-sheet in the target space $M$.
For compact world-sheets, this gives a contribution of
$\exp{ik\int \b}$ to the functional integral and so this will be a symmetry provided
$\frac {k}{2\p} \b$ represents an integral cohomology class.

Then the theory is invariant under $GL(d,\Z)$ and  integral shifts of $B$, in the sense that acting with these gives a physically equivalent theory.
For non-compact fibres, the situation is similar but 
the symmetries become the continuous symmetries of $GL(d,\R)$ and arbitrary constant shifts of $B$.

\section{Gauging the Ungaugable}\label{Ung}

Consider now the general case in which $(M,g,H)$ is invariant under the action of an abelian isometry group with $\cl_mH=0$  but in which the conditions for the gauging of the corresponding sigma-model are not necessarily satisfied, so that their consequences discussed in the previous sections also do not apply.
Then $\i_mH$ is closed but need not be exact.
Given a suitable good open cover $\{U_{\alpha}\}$ of   $M$, in 
  each patch $U_\a$ a one-form $v_m^\a $ can be found such that
  \begin{equation} \label{valis}
\i_mH =d v_m^\a
\end{equation}
In the overlap $U_{\alpha}\cap U_{\beta}$, the difference between the $v$'s must be closed and so exact, so that
\begin{equation} \label{vtrans}
v_m^\a
-v_m^\b= d \l^{\a\b}_m
\end{equation}
for some $ \l^{\a\b}$.
Then in triple overlaps,
$ \l^{\a\b}+ \l^{\b\g}+ \l^{\g\a}=c^{\a\b\g}$ for some constants $c^{\a\b\g}$.
If these constant cocyles vanish in all triple overlaps, then
 each $v_m$ is the  connection for some    line or circle bundle
 over $M$, and we now restrict ourselves to this case. This can be viewed as a restriction on the  group action on the B-field. There are then $d$ such connections $v_m$, so that they combine to form the connection for some bundle
  $\hat M$ over $M$ with $d$-dimensional fibres.  In the next section, it will be seen that this should be taken to be a torus bundle, with fibres $U(1)^d$.
  Choosing
fibre coordinates
$\hat X_m^\a $ over each  patch 
$U_\a$, 
with transition functions
\begin{equation}
\label{xhattrans}
\hat X_m^\a- \hat X_m^\b= - \l^{\a\b}_m
\end{equation}
then
\begin{equation} \label{vhat}
\hat v _m = d\hat  X_m^\a +  v_m^\a
\end{equation}
are globally defined 1-forms on $\hat M$ as
$\hat v^\a_m= \hat v_m^\b$ over $U_{\alpha}\cap U_{\beta}$.
In this section, it will be shown that the sigma model on $M$ can be lifted to a sigma-model on $\hat M$
and that under certain circumstances the isometries can be lifted to gaugable ones on $\hat M$, even if they were ungaugable on $M$.

Then $\hat M$ with coordinates 
$\hat X^I=( X^i, \hat  X_m)= (Y_\m , X^m, \hat  X_m)$ 
 is a bundle over $M$ with projection $\p : \hat M \to M$ with
$\p : (  X^i, \hat  X_m)\to ( X^i)$.
A metric $\hat g$ and closed 3-form $\hat H$ can be chosen on $\hat M$ with no $\hat  X_m $ components, i.e.
\begin{equation} \label{abc}
\hat g = \p ^* g, \qquad \hat H = \p ^* H
\end{equation}
where
$ \p ^*$ is the pull-back of the projection.
The pull-back will often be omitted in what follows, so that the above conditions will be abbreviated to  $\hat g=g, \hat H =H$. Then the only non-vanishing components of $\hat g _{IJ}$ are $\gij$ and $\partial/\pa \hat  X_m$ is a null vector,
while  the only non-vanishing components of $\hat H _{IJK}$ are $\hijk$.

It will be convenient to lift the Killing vectors $k_m$ on $M$ to
 vectors $\hat k_m$ on $\hat M$ that act on $\hat  X_m$ as well as $X^i$, so that
\begin{equation} \label{khatis}
\hat k_m=k_m + \Theta _{mn}  \frac {\pa} {\pa \hat  X_n}
\end{equation}
for some $\Theta _{mn}$. 
For $\hat k_m$ to be   vector fields on $\hat M$ requires, using (\ref{xhattrans}),  that 
$\Theta _{mn}$
have transition functions
\begin{equation}
\label{thtrans}
\Theta _{mn}^\a -\Theta _{mn}^\b=- \i _m d\l ^{\a\b}_n
\end{equation}

As $g,H$ are independent of $\hat  X$, the $\hat k_m$  are Killing vectors on $\hat M$:
\begin{equation}
\hat \cl _m \hat g=0, \qquad \hat \cl _m \hat H=0
\end{equation}
For any choice of $\Theta _{mn}$, there is an action generated by the Killing vector fields  $\hat k_m$
on the space $(\hat M,\hat g, \hat H)$
and we now turn to the question of whether this satisfies the conditions for gauging reviewed in section 2.
If $\hat \i_m$ denotes the interior product with $\hat k_m$,
then 
\begin{equation}
\hat \i_m\hat v_n = \i_m v_n +\Theta _{mn}
\end{equation}
If $\Theta _{mn} $ is chosen to be 
\begin{equation} \label{thiss}
\Theta _{mn} = B_{mn} - \i_m v_n
\end{equation}
for some antisymmetric $B _{mn}=-B _{nm}$, then
\begin{equation} \label{ivhat}
\hat \i_m\hat v_n +\hat \i_n\hat v_m
=0
\end{equation}
Further, as $dv= d \hat v$,
\begin{equation} \label{abc}
\hat \i_m \hat H = d \hat v_m
\end{equation}
Next, the Lie derivative of $\hat v $ with respect to $\hat k$ is
\begin{equation} \label{abc}
\hat \cl _m \hat v _n = \hat \i _m \hat \i _n \hat H +  d \hat \i_m\hat v_n =
  \i _m  \i _n H + d B_{mn}
\end{equation}
  so that if
  \begin{equation} \label{iihh}
  \i _m  \i _n H =-  d B_{mn}
\end{equation}
  then
  \begin{equation} \label{abc}
\hat \cl _m \hat v _n=0
\end{equation}
Then   $\Theta$  has the transition functions
(\ref{thtrans})
provided $B_{mn}$ are globally defined functions on $ M$, $B_{mn}^\a=B_{mn}^\b$,  and this together with  (\ref{iihh})
 implies that $\i _m  \i _n H $ is exact.

Finally,  consider the   algebra for the isometries generated by the $\hat k_m$.
  The Lie bracket is
  \begin{equation} \label{abc}
[\hat k_m, \hat k_n]=2 \cl _{[m} \Theta _{n]p}  \frac {\pa} {\pa \hat  X_p}
\end{equation}
  Using (\ref{iihh}) and $\cl _{m} B _{np} = \i_m d B_{np}$ since $B_{np}$ is a 0-form, one finds
  \begin{equation} \label{lieb}
 \cl _{m} B _{np} = -\i_m \i_n\i_p H
\end{equation}
  while (\ref{valis}) implies
  \begin{equation} \label{abc}
2 \cl _{[m} \i _{n]}v_p=-\i_m \i_n\i_p H
\end{equation}
  Then the Lie bracket  is 
  \begin{equation} \label{abc}
[\hat k_m, \hat k_n]=-(\i_m \i_n\i_p H)
  \frac {\pa} {\pa \hat  X_p}
\end{equation}
so that the algebra is abelian 
if
\begin{equation} \label{iabcH}
\i_m \i_n\i_p H
=0
\end{equation}
If this holds, then  (\ref{lieb})  implies that $B_{mn}$ is constant along the orbits of $ k$:
\begin{equation}
  \cl _m B_{np}=0
\end{equation}
Then $B_{np}$ are basic and  can be regarded as functions on $N$.

There are a further $d$ vector fields on $\hat M$ defined by
\begin{equation} \label{abc}
\ti k ^m = \frac {\pa}{\pa \hat  X_m}
\end{equation}
and   as $g,H$ are independent of  $ \hat  X_m$, these are Killing vectors preserving $H$. Then
$\hat M$ has $2d$ commuting Killing vectors $\hat k_m, \ti k^m$.
Assuming $G_{mn}= \hat g ( \hat k_m, \hat k_n)=g (k_m,k_n)$ is invertible,
the one forms 
\begin{equation}
\label{}
\hat \x ^m \equiv  G^{mn} \hat g _{IJ} \hat k ^I_n dX^J= \x^m
\end{equation}
are the same as $\x^m$.
The one-forms $\ti \x_m$ defined by 
\begin{equation}
\hat v_m =  \ti \x_m-B_{mn}  \x ^n 
\end{equation}
are horizontal with respect to   $\hat k_m$.

It is useful to choose local coordinates  $(\ti X^m, \ti X_m , \ti Y^\m)$ adapted to the $2d$ commuting  isometries, so that
\begin{equation}
\hat k _m = \frac {\pa}{\pa \ti  X^ m}, \qquad \ti k ^m = \frac {\pa}{\pa \ti  X_m}
\end{equation}
The required change of coordinates is
\begin{eqnarray}
\ti X^m &=& X^m \nonumber \\
\ti Y^\m &=& Y^\m  \nonumber\\
\ti X_m &=& \hat X_m+ f_m \label{coords}
\end{eqnarray}
where
$f_m(X^m,Y^\m)$ satisfies
\begin{equation}
\label{fdiff}
\frac{\pa f_m}{\pa X^n}=- \Theta _{nm}
\end{equation}
so that
\begin{equation}
d\ti X_m = d \hat X_m - \Theta_{nm} dX^n+ f_{m,\m}dY^\m
\end{equation}
The integrability condition $\pa _{[p} \Theta_{n]m}=0$ 
for (\ref{fdiff}) is satisfied as a result of (\ref{iabcH}).
Then in the coordinate system $(X^m, \ti X_m,Y^\m)$ many of the results derived in section 3 can be applied. In particular, 
\begin{equation}
\ti \x_m = d\ti X_m+ \ti A_m 
\end{equation}
where $\ti A_m= \ti A_{m\m }dY ^\m $
is a connection one-form 
that  is horizontal with respect to  $\hat k_m, \ti k^m$, and
\begin{equation}
\ti F_m = d\ti \x_m = d\ti A_m
\end{equation}
is also horizontal. 

Then the  geometry $(\hat M, \hat g, \hat H)$  with doubled fibres can be constructed provided the closed 2-form $\i _m H$ is the curvature for some line bundle. The
sigma-model on $(\hat M, \hat g, \hat H)$ has an abelian isometry symmetry
generated by the $\hat k_m$ which can be gauged precisely if 
the original geometry $(M,g,H)$ has as an isometry generated by the $k_m$ 
satisfying  the two conditions that (i)     $  \i _m  \i _n H$
 is exact, so that there are well-defined functions $B_{mn}$ on $M$ satisfying (\ref{iihh}), and (ii) $\i_m \i_n\i_p H
=0$.
 These are considerably weaker than the conditions needed for the isometry of 
 $(M,g,H)$ to be gaugable; 
 here $v_m^\a $ need not be globally defined, 
and is not required to satisfy either   $\cl _m v_n=0$ or
 $\i_mv_n = -\i_n v_m$.
A more general construction in which condition  (i) is relaxed will be discussed in later sections.

The gauged action is now obtained by inserting the appropriate hatted objects in (\ref{gag3}) or (\ref{gag2}).
The action  (\ref{gag3})  becomes
\begin{equation} \label{gaghat}
\hat S= \frac12 \int  _W    \gij  D  X^i \we *D X^j +   \int  _V   \left(  \frac13 \hijk   D X^i \we D X^j \we D  X^k  +  \cg ^m{} \we \hat v_{mI} D  \hat X^I \right)
\end{equation}
where
\begin{equation} \label{abc}
D_a \hat X^I =\pa _a  \hat X^I - C_a ^m \hat k_m^I 
\end{equation}
so that
\begin{equation} \label{abc}
D_a\hat  X_m= \pa _a\hat  X_m + \Theta _{mn} C^n_a
\end{equation}

The action  can be rewritten as
\begin{equation} \label{gaghat2}
\hat S = S_0 + \int _W \left(- C^m_a\hat  J^a_m + \ha C^m_aC^n_b\left[ G_{mn} \h^{ab} +B_{mn} \e ^{ab}\right]
\right)
\end{equation}
where
\begin{equation} \label{abc}
\hat  J^a_m= J^a_m 
- \e ^{ab} \pa _b \hat  X_m
\end{equation}
(Note that $\hat g_{IJ } \hat k^J_m d\hat X^I = \gij k^i_m dX^j$, and  $\hat \x^m = \x^m$ is dual to $\hat k_m$.)

 As before, shifting  the gauge fields  $C$ gives the action (\ref{fact}) plus 
\begin{equation} \label{abc}
S'=S_0- \int _W d^2\s \,  \hat J^-_m (E^{-1})^{mn} \hat J^+_n
\end{equation}
so that the original action
\begin{equation} \label{abc}
S_0=
\int _W d^2\s \,  \hat  \ce _{IJ} \pa _+ \hat X^I \pa _- \hat X^J
=
\int _W d^2\s \,  \ce _{ij} \pa _+ X^i \pa _- X^j
\end{equation}
is changed by replacing  $\hat  \ce _{IJ}$
with
\begin{equation} \label{abc}
 \hat \ce '_{IJ}= \hat \ce _{IJ}- ( \hat k_{mI}+  \hat v_{mI} )(E^{-1})^{mn}  ( \hat k_{mJ}- \hat v_{mJ} )
\end{equation}
which can be rewritten as
\begin{equation}\label{repoi}
\ce '_{IJ}=\ce _{IJ}- (E_{pm}  \x_{I}^p+ \ti \x_{mI} )(E^{-1})^{mn}  (E_{nq}  \xi_{J}^q-\ti \x _{nJ} )
 \end{equation}
 with symmetric and anti-symmetric parts
 \begin{eqnarray}\label{repoi2}
g' & = & g - G_{mn}   \x ^m \otimes   \x ^n + \ti G^{mn} \ti \x _m \otimes \ti \x _n \\
b' & = & b - \ti \x_m \we    \x ^m  -  \ha  B_{mn} \x ^m\we  \xi ^n +\ha \ti B^{mn}\ti \x_{m} \we \ti \x _{n}
\end{eqnarray}

\section{Global Structure and Large Gauge Transformations}

In the last section, it was seen that adding extra coordinates $\hat X$ 
enables one to overcome obstructions to gauge a wide class of sigma-models.
This involved replacing $v$ with $\hat v= d\hat X+v$ and the gauged action
  $\hat S$  (\ref{gaghat2}) differs from   (\ref{gag2})   by an extra term proportional to $\hat v - v$,
\begin{equation} \label{ext}
  \int  _W   C ^m \we    d\hat   X_m
\end{equation}
Suppose that the orbits of the $k_m$ are compact, so that $X^m$ are periodic coordinates on a torus.
Then the question arises as to whether the new coordinates are also periodic.
In \cite{Alvarez:1993qi}, it was argued that the invariance of the extra term in (\ref{ext}) under large gauge transformations requires that
$\hat X$ be periodic, However, the situation is complicated due to the fact that $\hat X$ is not invariant under the transformations generated by $\hat k$, and the action $S_0$ in  (\ref{gag2}) may not be invariant under large gauge transformations in general.
In this section, it will be shown that $\ti X_m$ are periodic coordinates for a torus dual to the $X^m$
torus. Note that from (\ref{coords}),   periodicity conditions for $\ti X$ are not consistent with periodicity conditions for $\hat X$ unless the components  of $\Theta _{mn} $ are rational numbers, and as $\Theta _{mn} $  varies continuously over $N$ this will not be the case in general.
WIth the coordinates $\ti X$ periodically identified, the orbits of the $\ti k^m$ are periodic and the space $\hat M$ is a torus bundle over $N$ with fibre $T^{2d}$.

\subsection{Simplified Form of Gauged Sigma-Models}

Consider the gauged sigma-model on $(M,g,H)$ discussed in sections 2,3.
As
\begin{equation} \label{abc}
H=\bar H +  \ti F_m \we   \x^m + dB
\end{equation}
where
\begin{equation}
B=\ha   B_{mn}
   \x ^m \we  \x ^n 
\end{equation}
    is a globally-defined 2-form, the pull-back  $\f^*B$ defines
    a WZ-term $\int_W \f^*B$ which can be gauged by minimal coupling.
    The gauged action is then the sum of the minimal coupling term
    \begin{equation} \label{abc}
S_{min}= \frac12 \int  _W    \gij  D  X^i \we *D  X^j +   B_{mn} \x^m_i\x ^n_j DX^i \we DX^j
\end{equation}
and a non-minimal term
\begin{equation}
  S_{non-min}=   \int  _V   \left(  \frac13 (H-dB)_{ijk}   D  X^i \we D  X^j \we D X^k  +  \cg ^m  \we  \bar \x_{mi} DX^i \right)
\end{equation}
which can be rewritten locally as
 \begin{equation} \label{abc}
 S_{non-min}=   
  \int _W  (b-B)+ C^m\we \bar \x_m = \int _W d^2\s \, \e ^{ab}\left( \ha (b-B)_{ij} \pa _a X^i \pa _b X^j+C^m_a \bar \x_{mi} \pa_b X^i\right)
  \end{equation}   
where $\bar \x$ is defined by (\ref{vbis})

For the sigma-model on $(\hat M, g,H)$ with the action of $\hat k$ gauged, similar formulae apply with
\begin{equation}\label{nmin3}
  S_{non-min}=   \int  _V   \left(  \frac13 (H-dB)_{ijk}   D  X^i \we D  X^j \we D X^k  +  \cg ^m  \we  \ti \x_{mi} DX^i \right)
\end{equation}
The corresponding two-dimensional action is
 \begin{equation} \label{nmin2}
 S_{non-min}=   
  \int _W  (b-B)+ C^m\we \ti \x_m   \end{equation}   

\subsection{Large Gauge Transformations and Global Structure}

A homology basis of one-cycles on $\hat M$    {} $(\g _n,\ti \g ^n, \g_A)$
can be chosen so that $\g _m$ is the one-cycle generated by $k_m$, $\ti \g ^m$  is the one-cycle generated by $\ti k^m$, and $\g_A$ are one-cycles on $N$.
Then  the periods are
\begin{equation}
\label{pedjfjhk}
\oint _{\g _n} \x^m = 2\p R_m \d ^m{}_n, \qquad \oint _{\ti \g ^n} \ti \x_m = 2\p \ti R_m \d _m{}^n
\end{equation}
for some $R_m, \ti R_m$, and
in the adapted coordinates this determines the periodicities
\begin{equation}
X^m \sim X^m + 2\pi R^m, \qquad \ti X_m \sim \ti X_m+ 2\pi \ti R_m
\end{equation}
From the form of the minimal couplings, for any 1-cycle $g$
on $W$, the Wilson line $\oint _g C$ transforms under a large gauge transformation $g:W \to U(1)^d$
with winding numbers $N^m$ ($m=1,...d$) around $\g$ as 
\begin{equation}
\oint _g C^m \to \oint _g C^m +
 2\p N^m R^m 
\end{equation}
 Then the
 change in the term $\int C^m\we \ti \x_m $ in the non-minimal   action (\ref{nmin2})
 will leave the functional integral invariant 
 provided the radii are inversely related, so that for each $m$
 \begin{equation}
 \label{quantcon}
2\p k R_m\ti R_m \in \Z
\end{equation}
 
The ambiguity in the three-dimensional form of the non-minimal term (\ref{nmin3}) for  two 3-manifolds  $V,V'$ with the same boundary $W$ is the integral over the compact 3-manifold 
$V-V'$
\begin{equation}
S_{non-min}(V)-S_{non-min}(V')= \ha \int  _{V-V'}  \cg ^m  \we  \ti \x_{mi} DX^i 
\end{equation}
The integral of $ \cg ^m$ over any   2-cycle $\G\in W$ is
\begin{equation}\label{quanty}
\int _\G \cg ^m = 2\p N R^m
\end{equation}
for some integer $N$. Then
the integral over the compact 3-manifold 
$V-V'$ will not affect the functional integral provided the same condition (\ref{quanty}) is satisfied.

Thus the torus generated by the $\ti k$ with coordinates $\ti X_m$ is dual to the
torus generated by the $ k$ with coordinates $X^m$, with inversely related periodicities  (\ref{quanty}).
A convenient choice is to take $R_m=1, \ti R= 1/(2\p k)$ for all $m$.
For each $m$, $X^m/R^m$ has period $2\p$ and $C^m/R^m$ is conventionally normalised, so that
for any   2-cycle $\G\in N$
\begin{equation}\label{quantcy}
\int _\G \F ^m = 2\p N R^m
\end{equation}
for some integer $N$, so that $( 2\p   R^m)^{-1}[F^m]$ represents an integral cohomology class for each $m$.
The condition that $(k/2\p) [H]$ is an integral cohomlogy class implies from (\ref{hiso}) that $(k/2\p) [\ti F_m\we \x ^m]$
should also be an integral cohomology class. Using (\ref{pedjfjhk}), this implies that
$kR_m [\ti F_m] $ be integral, and using (\ref{quantcon}) this implies
 that
  $( 2\p   \ti R_m)^{-1}[\ti F_m]$ is integral, so that the topology 
 is partially characterised by
  $d$ Chern-classes $( 2\p   R^m)^{-1}[F^m]$ and $d$ dual Chern classes or $H$-classes
  $( 2\p   \ti R_m)^{-1}[\ti F_m]$ in $H^2(N,\Z)$.

Consider now the integration over $\ti X_m$ for arbitrary $W$, following  \cite{Rocek:1991ps},\cite{Giveon:1991jj},\cite{Alvarez:1993qi}.
On a general Riemann surface $W$, $\ti X_m(\s)$ can be written in terms of 
a function $x_m(\s) $ and a winding term, so that
\begin{equation}
\label{dfgssfdg}
d\ti X_m(\s)= d x_m (\s)+ \sum_r 2\p N_m^r \ti R_m  \w _r(\s)
\end{equation}
 where 
$\{ \w _r \}$
is a basis of harmonic 1-forms on $W$ (normalised to have integral periods) and  $N_m^r$ are integers. 
Then the only dependence on $\ti X $ of (\ref{nmin2})  
is through the term  $C^m\we d\ti X_m$, so that using (\ref{dfgssfdg}), the functional  integral 
over $\ti X_m$ becomes a functional integral over $x_m $ and a sum over the integers $N_m^r$.
The $x_m $  are lagrange multipliers imposing the constraint 
$\cg ^m=0$, so that $C^m$ are flat connections, while the
sum over the integers $N_m^r$ imposes the constraint that the
Wilson lines $\oint C$ all vanish, so that the connection
$C$ is pure gauge.
Then a suitable gauge choice is $C=0$, in which case the ungauged model is recovered.

 \section{T-Duality} \label{TDuality}

\subsection{T-Dualising on $d$ Circles}

If $X^m$ are coordinates on a torus, the $\ti X_m$ are coordinates on the dual torus.
$M$ is a $T^d$ bundle over $N$, and 
$\hat M$ is  a torus bundle over $M$ and so a $T^{2d}$ bundle over $N$.
With these periodicities, it was seen in the last section that 
$\ti X_m$ is a lagrange multiplier imposing the condition that $C$ is pure gauge, and
so 
can be set to zero by a gauge choice, and the  ungauged model  on $(M,g,H)$ is recovered.
Then the gauged model on $(\hat M,g,H)$ (\ref{gaghat})  or (\ref{gaghat2}) is equivalent 
to the ungauged model  on $(M,g,H)$ for any $W$.
However, one can instead integrate out the gauge fields $C$ to get a sigma model with
  geometry $(\hat M, g',H')$ given by (\ref{repoi}) or (\ref{repoi2}). This still has the local gauge symmetry (\ref{trans}),
and taking the quotient by the isometry group generated by the $\hat k_m$ gives
a sigma-model on $\ti M$, the space of orbits, with metric $\ti g = g'$ and 3-form $\ti H= H'$.
Then the sigma-model on
 $(\ti M,\ti g, \ti H)$ is equivalent to that on $(M,g,H)$  as they define equivalent quantum theories, since 
  the   functional integrals are related by different gauge choices for the master sigma-model on
  $\hat M$.
  The projection from the model on $\hat M$ to that on $\ti M$ can be thought of as a gauge-fixing 
of the isometry symmetry by setting the $X^m$ to constants locally.

The formulae from section 3 can be immediately applied to this case of the gauging of the sigma-model on $\hat M$, with the replacement $\hat \x \to \ti \x$.
For $d=1$, from (\ref{fdhkasf}), the metric $g$ on $M$ and dual metric $\ti g$ on $\ti M$ are
\begin{eqnarray}
g &=& \bar g + G \, \x \otimes \x  \\
\ti g &=&\bar g  +G^{-1} \, \ti \x \otimes \ti \x
\end{eqnarray}
while the 3-form $H$ and dual 3-form $\ti H$ are, using (\ref{kjdflkadjflk}), 
\begin{eqnarray}
H & = &  \bar H+\x  \we  \ti  F
\\
\ti H & = &  \bar H+\ti  \x  \we   F
\end{eqnarray}
and $\bar H$ is a 3-form  satisfying
\begin{equation}
d\bar H= - F\we \ti F
\end{equation}
where
\begin{equation}
F=d\x, \qquad \ti F = d \ti \x
\end{equation}
There is a  Killing vector  $k$ on $M$ dual to $\x$, with
$g(k,V)=G\x(V)$ for any vector field $V$, 
 and a  Killing vector  $\ti k$ on $\ti M$ dual to
$\ti \x$. The forms $\bar H,F,\ti F$ are basic with respect to $k$ on $M$ and with respect to $\ti k$ on $\ti M$, so
can be viewed as forms on $N$.
These transformations agree with those found by Buscher locally, but are given in terms of globally defined 
objects.
In local coordinates adapted to the Killing vectors,
\begin{equation} \label{abc}
k = \frac {\pa} {\pa X}, \qquad \ti k = \frac {\pa} {\pa \ti X}
\end{equation}
and
\begin{eqnarray}
\x & = & dX+A \\
\ti \x & = & d\ti X + \ti A 
\end{eqnarray}

There is a straightforward generalisation to T-dualising on $d$ circles.
Using (\ref{kdjjsfh}),(\ref{kdjjsfh2}) with  $\hat \x \to \ti \x$,   the original geometry $(M,g,H)$ and the dual geometry $(\ti M,\ti g, \ti H)$ are
given by
 \begin{eqnarray}
 \label{tgd}
g &=& \bar g + G _{mn}\x ^m\otimes \x^n  \\
\ti g &=&\bar g  +\ti G^{mn} \, \ti \x _m\otimes \ti \x _n
\end{eqnarray}
and
\begin{eqnarray}
\label{thd}
H&=&\bar H +  \ti F_m \we   \x^m + dB
\\
\ti H& =& \bar H + \ti \x _m \we  F^m + d\ti B\end{eqnarray}
Here $E=G+B$ and
\begin{equation}
\ti G^{mn}= (E^{-1})^{(mn)}, \qquad \ti B^{mn}= (E^{-1})^{[mn]}
\end{equation}
while
\begin{equation}
B=\ha   B_{mn}
   \x ^m \we  \x ^n , \qquad \ti B=\ha \ti   B^{mn}
   \ti \x _m \we  \ti \x _n 
\end{equation}
   and 
   \begin{equation}
F^m=d\x ^m, \qquad \ti F _m = d \ti \x_m
\end{equation}
while
   $\bar H$ satisfies
   \begin{equation}
d\bar H= - \ti F_m \we F^m
\end{equation}
 There are  $d$ Killing vectors $k_m$ on $M$ dual to $\x^m$ 
    and $d$  Killing vectors  $\ti k^m$ on $\ti M$ dual to
$\ti \x$ and the forms $\bar H,F^m,\ti F_m$ are basic with respect to $k_m$ on $M$ and with respect to $\ti k^m$ on $\ti M$, so
can be viewed as forms on $N$.
In adapted local coordinates
\begin{equation} \label{abc}
k_m = \frac {\pa} {\pa X^m}, \qquad \ti k^m = \frac {\pa} {\pa \ti X_m}
\end{equation}
and
\begin{eqnarray}
\x ^m & = & dX^m+A ^m \\
\ti \x _m& = & d\ti X _m+ \ti A _m
\end{eqnarray}

Thus the effect of T-duality is to change
the bundle $M$ over $N$ with fibres generated by $k_m$
to the dual bundle
 $\ti M$ over $N$ with fibres generated by $\ti k^m$
while the geometries are interchanged by
\begin{equation}\label{x-x}
\x^m \leftrightarrow \ti \x_m
\end{equation}
and   
\begin{equation}\label{e-e}
E  \leftrightarrow \ti E \equiv  E^{-1}
\end{equation}
This implies that the 1st Chern classes are interchanged with the  $H$-classes, which are the dual  1st Chern classes
\begin{equation}\label{x-x}
[F^m] \leftrightarrow [ \ti F_m]
\end{equation}

\subsection{The Action of $O(d,d;\Z)$}

The geometry of a $T^d$ bundle $(M,g,H)$ with $d$ Killing vectors satisfying the conditions of section 4 is specified by
the base geometry on $N$ specified  by $\bar g, \bar b$, the $2d$ vector potentials $A^m, \ti A_m$, and 
the scalars $G_{mn}, B_{mn}$. The base geometry is then
  $(N,\bar g, \bar H)$ with $\bar H $   given by (\ref{rtrhrthrmh}).
There is a natural action of $GL(d,\R)$ on $A^m, \ti A_m$, and $G_{mn}, B_{mn}$ and it was seen that the transformation under $GL(d,\Z)$ or under integral shifts of the $B$ field
  takes the geometry to one defining the same quantum field theory.
  The T-duality transformation discussed in the 
  last subsection  dualises in $d$ circles to obtain a dual geometry $(\ti M,\ti g,\ti H)$ 
  defining the same quantum theory.
  Such a 
T-duality transformation can be applied to any $d'\le d$ of the circles, giving further dual geometries.
The group generated by $GL(d,\Z)$, integral $B$-shifts and the T-dualities on any $d'\le d$ circles is $O(d,d;\Z)$.
The action of  $O(d,d;\Z)$ is given as follows.

Consider an  $O(d,d)$ transformation
by
 \begin{equation}\label{oddm}
h=\left(
\begin{array}{cc}
a & b\\
c & d \end{array}\right),
  \end{equation}
where $a,b,c,d$ are $d\times d$ matrices.
This preserves the indefinite metric  \begin{equation}
L=\left(\begin{array}{cc}
0 & \1 \\
\1 & 0
\end{array}\right)
\label{liss}  \end{equation}
 so that
\begin{equation}
h^tLh=L\;\; \Rightarrow \;\;a^t c+c^t a=0,\;\;\;b^t d+d^t
b=0,\;\;\;\ a^t d+c^t b=\1 . \label{abc}  \end{equation} The
transformation rules for $E$
 give  the
non-linear   transformation of $E$ under a T-duality
transformation $h \in O(n,n )$ \cite{GMR},\cite{Giveon:1991jj},\cite{Giveon:1994fu}
\begin{equation}
E'  = (aE+b)(cE+d)^{-1}. \label{tetrans}  \end{equation} 
The  $2d$ 1-forms $\x,\ti \x$ 
combine into a $2d$ vector of 1-forms
\begin{equation}
  \Xi =
\left( \begin{array}{c} \x^m\\ \ti \x_m  \end{array}\right)
\label{xicol}  \end{equation}
transforming as a vector under $O(d,d)$:
\begin{equation}
\label{xxitrans}
\Xi \to \Xi ' = h^{-1} \Xi
\end{equation}
The group $O(d,d,\Z)$ consists of matrices (\ref{oddm}) with integral entries.

The $GL(d;\Z)$ subgroup 
is 
 \begin{equation}
h_L=\left(\begin{array}{cc}
\ti L & 0 \\
0 & L
\end{array}\right)
\label{gL}  \end{equation}
where
$L_m{}^n \in GL(d;\Z)$ and $\ti L = (L^t)^{-1}$.
The subgroup of B-shifts $B \to B+\b$ is through matrices of the form
  \begin{equation}
h_\b= \left(\begin{array}{cc}
\1 & \b \\
0 & \1
\end{array}\right)
\label{gB}  \end{equation}
for integral $\b$.
The subgroup $\G(\Z)$ of matrices of the form
   \begin{equation}
h_\G= \left(\begin{array}{cc}
\ti L & \b \\
0 & L
\end{array}\right)
\label{gG}  \end{equation}
plays  an important role, and will be referred to as the geometric subgroup.

The transformation T-dualising
in all $d$ circles is
  \begin{equation}
h_T=\left(\begin{array}{cc}
0 & \1 \\
\1 & 0
\end{array}\right)
\label{gT}  \end{equation}

In adapted coordinates
\begin{equation}
\label{ }
\Xi=d  \cx + \ca
\end{equation}
where, introducing $O(d,d)$ vector indices $M=1,...,2d$,
\begin{equation}
\label{xcol}
  \ca ^M =
\left( \begin{array}{c} A^m\\ \ti A_m  \end{array}\right),
\qquad
 \cx ^M=
\left( \begin{array}{c} X^m\\ \ti X_m  \end{array}\right)
 \end{equation}
also transform  as a vector under $O(d,d)$:
\begin{equation}
\label{atranss}
\ca \to \ca' = h^{-1} \ca, \qquad \cx \to \cx' = h^{-1} \cx
\end{equation}
Then the $\cx$ are fibre coordinates for a $T^{2d}$ bundle over $N$ with connection 1-forms $\ca$ \cite{Hull:2004in}.
There are $2d$  field strengths $\cf = d \ca$, and the corresponding 1st Chern classes
$[\cf ]$ transform as 
\begin{equation}
\label{ftranss}
[\cf  ]\to [\cf' ] = h^{-1}[ \cf ]
\end{equation}
Then the $d$ Chern classes  and  the $d$  $H$-classes
fit into a $2d$-dimensional representation and are mixed together under the action of $O(d,d;\Z)$.

\section{Torus Fibrations}

\subsection{Local Killing Vectors}

For string theory on a space that is a $K$ bundle,
i.e. a bundle   whose fibres are some space $K$,  
there are general arguments  \cite{Vafa:1995gm} that any duality that applies to string theory on $K$ (e.g. mirror symmetry 
if $K$ is Calabi-Yau, or T or U dualities if $K$ is a torus)
 can be applied fibrewise, giving a fibration by  a dual string theory on a space whose fibres are the dual space $\ti K$.
 In the present context, this implies that it should be possible to apply T-duality 
to any space with a $T^d$ fibration.
However, the arguments discussed so far have been based on the case where there is an isometry group
generated by globally defined Killing vector fields.
In this section, these will be generalised to general torus fibrations, which do not have globally defined Killing vector fields. The aim of this section is to give a direct proof that T-duality can be applied fibrewise, and to examine whether there can be obstructions to fibrewise T-duality.

In general, a $T^d$ bundle over $N$ can have $GL(d,\Z)$ monodromy  around each  1-cycle $\g$ in $N$,
with the fibres twisted by a large diffeomorphism on $T^d$, so that if $k_m$ are the  vector fields generating periodic motions  along the $T^d$
 fibres, 
then continuing $k_m$ round $\g$ brings it back to a linear combination $L_m{}^n(\g)k_n$ of the vectors $k_m$. Then although there are locally defined Killing vectors, they do not extend to global Killing vector fields -- if one tries to analytically continue a solution of Killing's equation 
to the whole space,   non-trivial monodromy would imply that the vector field is multi-valued.

Suppose then that in each patch $U_\a$ of $M$ there are $d$ Killing vector fields
$k_m^\a$ such that 
$\cl _m g=0$, $\cl _m H=0$ in  $U_\a$, and that in each
overlap $U_{\alpha}\cap U_{\beta}$
\begin{equation}
\label{kpatch}
k_m^\a = (  L_{\a\b})_m{}^n k_m^\b
\end{equation}
for some matrix  $(  L_{\a\b})_m{}^n $
in $GL(d,\Z)$.\footnote{ The indices $\a,\b$ indicate the patch in which the corresponding function has support, while the composite index $\a\b$ indicates a function in the overlap $U_\a\cap U_\b$.
 There is no significance here as to whether they are subscripts or superscripts.}
It then follows that   objects constructed from $k_m$ and carrying indices $m,n...$ now have $GL(d,\Z)$ transition functions. For example, from their definitions it follows that $G, \x$ have transition functions
\begin{equation}\label{sdflkhsd}
G_\a=  LG_\b L^t,  \qquad \x_\a=\ti L\x_\b
\end{equation}
where $L= L_{\a\b}$ and $\ti L ^m{}_n$ is given by $\ti L= (L^t)^{-1}$.
Objects such as $G, \x$ carrying indices $m,n...$ whose  transition functions are just the $GL(d,\Z)$ transformation in the appropriate representation will be referred to as tensors.

If $X^m_\a$ are coordinates adapted to $k_m^\a$, so that $k_m^\a=\pa/\pa X^m_\a$, then
\begin{equation}\label{transxi}
\x_\a ^m = d X_\a^m +A^m_\a
\end{equation}
and 
\begin{equation}\label{transa}
A^m_\a =(\ti L _{\a\b})^m{}_n A^n_\b + d \r ^m_{\a\b}  
\end{equation}
and
\begin{equation}\label{transx}
X^m_\a= (\ti L_{\a\b} )^m{}_n X^n_\b -  \r ^m_{\a\b}  
\end{equation}
for some $ \r ^m_{\a\b}  $.
These are not tensorial patching conditions. The transition functions for the coordinates $X$ are an affine transformation, so such a bundle is sometimes referred to as an affine bundle.
Here $ \r ^m_{\a\b} $ satisfies $\i_m d \r ^n =0$, and so is a function on the base $N$.
The transition functions 
then act 
  by a large diffeomorphism of the torus together with a 
translation of the $X^m$, and so define an affine torus bundle rather than a principle one.

Next, as $\i _m^\a H = (L_{\a\b})_m{}^n \i_n^\b H$
(where $\i _m^\a$ is the interior product with $k_m^\a$)
\begin{equation}
dv^\a_m =(L_{\a\b})_m{}^n dv^\b_n\end{equation}
so that (\ref{vtrans}) is replaced with
\begin{equation} \label{abc}
v_m^\a
-(L_{\a\b})_m{}^nv_n^\b= d \l^{\a\b}_m
\end{equation}
Then \begin{equation} \label{hatvis}
\hat v _m ^\a= d\hat X_m^\a +  v_m^\a
\end{equation}
will have covariant transition functions 
\begin{equation}
\label{hatvtrans}
\hat v_\a=L \hat  v_\b
\end{equation}
provided
\begin{equation}
\label{hatvtransl}
\hat X_m^\a = (L_{\a\b})_m{}^n\hat X_n^\b  - \l^{\a\b}_m
\end{equation}
The transition functions for $\Theta$ are
now 
\begin{equation}
\label{thtransa}
\Theta _{mn}^\a - L_m{}^p L_n {}^q\Theta _{pq}^\b=- \i _m d\l ^{\a\b}_n
\end{equation}

The one-forms $\ti \x_m^\a $ defined by 
\begin{equation}
\hat v_m ^\a  =  \ti \x_m-B_{mn}^\a   \x ^n_\a 
\end{equation}
will be tensorial, with
\begin{equation}
\label{}
\ti \x _m ^\a = ( L _{\a\b})_m{}^n\ti \x _n ^\b  
\end{equation}
provided the $B_{mn}$ are tensorial, $B^\a = LB^\b L^t$.
This condition will be assumed to be the case in this section, but more general transition functions for $B_{mn}$ will be discussed in section 8.
The 1-forms $\ti \x$
take the form
\begin{equation}
\ti \x _m ^\a = d\ti X_m ^\a +\ti A_m ^\a
\end{equation}
after the change of coordinates (\ref{coords}),(\ref{fdiff}) in each patch $U_\a$.
From (\ref{hatvtransl}),(\ref{thtransa}),(\ref{coords}),(\ref{fdiff}), it follows that
$\pa _p (\ti X_m^ \a- ( L_{\a\b} )_m{}^n X_n^\b )=0$
so that there are functions $\ti  \r _m^ {\a\b} $ on $N$ such that 
the patching conditions are
\begin{equation}
\ti A_m^\a =( L _{\a\b})_m{}^n \ti A_n^\b + d \ti \r _m^ {\a\b}  
\end{equation}
and
\begin{equation}\label{xitrans}
\ti X_m^ \a= ( L_{\a\b} )_m{}^n \ti X_n^\b -  \ti \r _m^{\a\b}  
\end{equation}
Then the bundle $\ti M$ over $N$ with fibres $\ti X$ and connection $\ti A$ is   a dual affine bundle.

If $M$ is a $T^d$ bundle over $N$, one can choose  a cover  for $M$ of sets $U_\a  \simeq \bar U_\a \times T^d$ where $\bar U_\a $ is an open cover of $N$. The transition functions discussed above are then all functions on intersections  $\bar U_{\alpha}\cap \bar U_{\beta}$ in $N$.

\subsection{Symmetries of Torus Fibrations and their Gauging}

In this section, geometries $(M,g,H)$ that are torus fibrations with local Killing vectors with transition functions (\ref{kpatch}) will be considered.
The formal  symmetries of the sigma-model on   $(M,g,H)$ that are associated with such local Killing vectors  will be discussed and their gauging
analysed.
This will then be used to discuss the symmetries and gauging  of the space $(\hat M,g,H)$ with doubled fibres and their implications for T-duality
in the following subsection.

A sigma-model configuration is a map $\f:W\to M$.
For a given   map  $\f:W\to M$, it is convenient to choose an open cover $W_{(\a,r)}$ 
(labelled by $\a$ and an extra   index $r$)
of $W$
such that
$\f (W_{(\a,r)}) \subset  U_\a$.
Such a cover can be constructed as follows. 
The map $\f$ can be combined with the bundle projection $\p :M \to N$ to define a map $\p \circ\f:W\to N$. Let
$\ti U_\a =\bar U_\a \cap ( \p \circ\f(W))$, so that $\{ u_\a \}$ with $u_\a = \f^{-1} \circ \p^{-1} \ti U_\a $ is a cover of $W$, with $\f (u_\a) \subseteq U_\a$.  For some $\a$, $u_\a$ may be the empty set.
Next, a good cover $\{ W_{(\a,r)} \}$ is chosen for 
each $u_\a$, $u_\a = \cup _r W_{(\a,r)}$ with contractible $W_{(\a,r)}$, and 
$W= \cup _{\a,r} W_{(\a,r)}$.

Then for $\s \in W_{(\a,r)}$, $\f(\s) \in U_\a$ and the coordinates $X^i_\a$ can be used.
Using $X^i_\a$  for $\s \in W_{(\a,r)}$ and  $X^i_\b$  for $\s \in W_{(\b,s)}$, for
$\s \in W_{(\a,r)}\cap  W_{(\b,s)}$, the transition functions following from (\ref{transx}) are
\begin{equation}\label{transxr}
X^m_\a(\s_{(\a,r)}) = (\ti L_{\a\b} )^m{}_n X^n_\b (\s_{(\b,s)}) -  \r ^m_{\a\b}  (\s_{(\b,s)}) 
\end{equation}
and the transition functions do not depend on $r,s$ (i.e. they are functions on $u_\a$).

 Consider the transformation of $X_\a(\s)$ for $\s$ in the patch $ W_{(\a,r)}$ given by
\begin{equation}
\label{xloc}
\d X^m _\a = \a^m_{(\a,r)} k_m^\a (X(\s))
\end{equation}
where the parameter
$\a^m_{(\a,r)}(\s)$ is a function on $ W_{(\a,r)}$.
As the patch $U_\a  \simeq \bar U_\a \times T^d$
  in $M$ contains the entire orbit of the each $k_m$, 
$X_\a +\d X_\a$ remains in $U_\a$ for each $\s \in W_{(\a,r)}$.
Consistency with (\ref{transxi}),(\ref{kpatch}) requires that, for $\s \in W_{(\a,r)}\cap  W_{(\b,s)}$,
 the parameters patch together according to
\begin{equation}
\label{apatchr}
(\a_{(\a,r)}) ^m= (\ti L_{\a\b} )^m{}_n ( \a _{(\b,s)})^n
\end{equation}

As the  transition functions (\ref{transxr})(\ref{apatchr}) do not depend on $r,s$, it follows that $X, \a$   are functions on $u_\a$
and for some purposes it is useful to use the cover $\{ u_\a \} $ and write the transition functions
for $X_\a(\s), \a_\a(\s)$ for $\s$ in $u_\a \cap u_{\b}
$
as 
\begin{equation}
\label{apatch}
\a_\a = \ti L \a _\b, \qquad X^m_\a= (\ti L_{\a\b} )^m{}_n X^n_\b -  \r ^m_{\a\b}  
\end{equation}
Note that the cover $\{ u_\a \}$ is not a good cover in general -- e.g. for the constant map $\f: W \to X_0 \in M$ 
of the whole world-sheet to a point $X_0\in U_{\a_0}$ for some patch $U_{\a_0}$, the corresponding patch $u_{\a_0}=W$ is the whole of $W$, and so this will not be contractible unless $W$ is.
For a rigid symmetry with constant $\a$,   a different constant parameter $\a_\b $ is needed in general for each patch $u_\b$, related by (\ref{apatch}).  The parameters are sections of a bundle, and in general
this has constant local sections, but not constant global sections.

Consider first the special case in which   $b$ is a tensor field with vanishing Lie derivative with respect to the vector fields $k_m$, so that the gauging is through minimal coupling, and $v_m = - \i _m b$. 
Defining $L^\a =L\vert _ {u_\a}$, the restriction of the ungauged sigma-model lagrangian $L(X(\s))$ to
$\s \in {u_\a}$, then  the coordinates $X_\a$ can be used and 
\begin{equation}
\label{lungage}
L^\a =\ha \gij d X^i \we  * dX^j + \ha \bij d X^i \we   dX^j
\end{equation}
where $X^i=X_\a^i$.
This extends to  a globally-defined lagrangian as
\begin{equation}
\label{ltrans}
L^\a=L^\b  \qquad {\rm in } ~ u_\a \cap u_b
\end{equation}

The transformation (\ref{xloc}) with constant $\a_\a$
is     a rigid symmetry of the lagrangian $L^\a$ for $\s \in u_\a$, and the question arises as to whether this extends to a symmetry of the full lagrangian on $W$. 
This will be the case if different constant parameters are chosen in each patch $u_\a \subset W$, with the
 transition functions (\ref{apatch}).
As the patching conditions for the parameters depend on the  choice of open sets $\{ u_\a \}$, and this in turn depends on a reference sigma-model map $\f : W \to M$, this is not  a proper rigid symmetry, but it is a formal invariance of the theory.

The transformation (\ref{xloc}) is a rigid symmetry of the lagrangian  $L^\a$ on $u_\a$
and this can be gauged by introducing 
the minimal coupling
\begin{equation}
\label{dxloc}
D _a X^i_\a = \pa _a X^m_\a - C^m_\a k^i_{\a m}
\end{equation}
where the connection one-forms $C_\a$ on $u_\a$ transform as 
\begin{equation}
\label{cloc}
\d C^m_{ a} = \pa _a \a ^m
\end{equation}
The minimal coupling gives the gauged lagrangian 
\begin{equation}
\label{lgage}
L^\a =\ha \gij D X^i \we  * DX^j + \ha \bij D X^i \we   DX^j
\end{equation}
where $X^i= X^i_\a$, $C = C_\a$  and this is invariant under the  local transformations (\ref{xloc}), (\ref{cloc}) on $u_\a$.
This can be done in   each patch, with a gauge field $C_\a(\s)$ for $\s \in u_\a$ in each patch.

These local gauged lagrangians will patch together to give a gauged lagrangian on $M$ that can be integrated over $W$ if (\ref{ltrans})
holds. 
Using (\ref{apatch}), this requires that the 1-forms $(C_\a)^m_a d\s ^a$ have   transition functions 
\begin{equation}
\label{ctrans}
C_{\a }  =(\ti L_{\a\b})  C _{\b }  -d \r   _{\a\b}
\end{equation}
where $d \r   _{\a\b}
$ is the pull-back
$d \r   _{\a\b}=d\s ^a \pa _a \r ^m _{\a\b}(X(\s))$.
The $C_{\a } $ are  1-forms on $u_\a$, so that 
if one had introduced $C_{(\a r)}$ on 
$W_{(\a,r)}$, then on the overlap
$W_{(\a,r)}\cap  W_{(\a, s)}$  the 1-form is continuous 
$C_{(\a ,r)}=  C_{(\a , s)}$, and the full form of the transition functions could be written
\begin{equation}
\label{ctransr}
C_{(\a,r)} =(\ti L_{\a\b})  C _{{(\b,s)}} -d \r   _{\a\b}
\end{equation}
and do not depend on $r,s$.
Comparing with (\ref{transa}), $C^m_\a$ has the same transition functions as the pull-back $-A^m_{\a i}\pa X^i_{\a a} d\s ^a$ of $-A$, so that $C$ is the connection of a bundle over  $W$
which  is the pull-back of the bundle 
$M$ over $N$  with connection $-A$.

As before, it is useful to write
\begin{equation}
\label{cfii}
C^m_{\a a} =\ti C^m_{\a a}  + \F^m_{\a a} 
\end{equation}
where
\begin{equation}
\label{ctiis}
\ti C_- =  (\x -E^{-1}\bar \x ) \pa _- X, \qquad \ti C_+ = (\x +(E^t)^{-1}\bar \x ) \pa _+ X
\end{equation}
The field equation from varying $C$ is $C=\ti C$ or, equivalently, $\F=0$.
The $\ti C$ is a pull-back connection, with transformation rules
 \begin{equation}
\ti C^m_{\a a}  =(\ti L_{\a\b})^m{}_n \ti C^n_{\b a}  -  \pa _a \r ^m _{\a\b}(X(\s))
\end{equation}
 so that $\F$ is a vector field with covariant transition functions
 \begin{equation}
 \label{fftrans}
\F^m_{\a a}  =(\ti L_{\a\b})^m{}_n \F^n_{\b a}   
\end{equation}
 in $u_\a \cap u_\b$.
 Any choice of $\F$ (e.g. $\F=0$) with these transition functions will give a $C$ with transition functions (\ref{ctrans}).

Then for each patch $u_\a$ there is a lagrangian $L^\a$ that is invariant under the local transformations 
(\ref{xloc}), (\ref{cloc}). Further,  if the gauge field  $C$ is a connection on the pull-back bundle, i.e. if it has
   transition functions (\ref{ctrans})
(or equivalently $C=\ti C + \F$ for any  $\F$ with transition functions (\ref{fftrans})),
then $L^\a=L^\b $ in $ u_\a \cap u_\b$
and the lagrangian
 is well-defined on $W$ and invariant 
under (\ref{xloc}), (\ref{cloc}) provided the local parameters
patch according to (\ref{apatch}).
The parameters $\a$ are 
local sections of a bundle with $GL(d,\Z)$ transition functions, and for non-trivial bundles, there will be no global constant section, and hence no global limit of the gauge symmetry with constant parameters. 
This bundle  is characterised by its $GL(d,\Z)$ monodromies around 1-cycles, and so 
can only  be trivial   if these monodromies are all trivial.
The best one can do in general is to find  constant local sections, with the $\a$ constant in each patch, but with the 
constants in different patches related by (\ref{apatch}). 

This can now be generalised to the case in which $b$ is not globally defined, but $H$ is invariant.
The gauging of the kinetic term involving the metric is as above.
The map $\f:W\to M$ extends to a map $\f:V\to M$ where $V$ is a 3-manifold with boundary $W$
and for any such map choose a   cover
$\{ V_\a \}$ of $V$ with $\p \circ\f(V_\a) \subset \bar U_\a$.
Then one can define the lagrangian on $V_\a$
\begin{equation} \label{gag4}
L^\a_{WZ}=     \frac13 \hijk   D  X^i \we D  X^j \we D X^k  +  \cg ^m  \we  v_{mi} DX^i  
\end{equation}
with $X=X_\a$. 
Assuming $\bar U_\a$ is contractible, there   
 are 1-forms
$v^\a_m$ in $U_\a$ such that 
$dv_m = \i _m H$. The $v_m$ are determined up to the addition of  exact forms, and the lagrangian 
$L^\a_{WZ}$
  is gauge invariant provided  the $v_m$ can be chosen so that
 $\cl _m v_n=0$ and $\i_{(m}v_{n)}=0$.
 These patch to give a well-defined action provided $L^\a_{WZ}=L^\b_{WZ} $ in $ V_\a \cap V_\b$, and this requires that the $v$ are tensorial:
 \begin{equation}
\label{vtrans}
v_\a = Lv_\b
\end{equation}
These give the generalisation of the conditions for gauging a Wess-Zumino term to the case of locally-defined Killing vectors.
The  connection has the same properties as above, and is given by 
(\ref{cfii}),(\ref{ctiis})  for any $\F$ 
with the transition functions (\ref{fftrans}).

\subsection{T-Duality for Torus Fibrations}

Suppose $M$ has $d$ locally-defined Killing vectors with transition functions (\ref{kpatch}).
If $\i_m \i _n \i _p H=0$, then over each patch $\bar U_\a$ in $N$ the construction of section 4 can be repeated to
give a patch $U_\a \simeq \bar U_\a \times T^{2d}$
with coordinates $(Y_\a ,X_\a , \ti X_{\a })$. This allows the construction of a space
 $\hat M$ which is a $T^{2d}$ bundle over $N$  
that has fibre coordinates $\cx_\a$ with
\begin{equation}
 \cx =
\left( \begin{array}{c} X^m\\ \ti X_m  \end{array}\right)
\label{abc}  \end{equation}
and patching conditions (\ref{transx}),(\ref{xitrans}).
The one-forms $\hat v$ defined by 
(\ref{hatvis})  are tensorial, with transition functions
(\ref{hatvtrans}).
There is   a $B^\a _{mn}$ and  vector fields $\hat k^\a_m$  in $U_\a$
such that the conditions for gauging are satisfied in $U_\a$, so that
a gauged lagrangian $L^\a$ can be constructed on $u_\a$ (or $V_\a$ for the WZ-term).

  The vector fields $\hat k^\a_m$ have the same tensorial transition functions as 
 $ k^\a_m$, $\hat k ^\a = L \hat k ^\b$
 provided the $B_{mn}$
given by 
 $ B_{mn} =\Theta _{mn} + \i_m v_n
$ are tensorial
 \begin{equation}
\label{btrnnn}
B^\a _{mn}=L_m{}^p L_n {}^q B^\b _{pq}
\end{equation}

Then in each patch 
there are torus moduli  $E^\a _{mn}=G^\a _{mn}+B^\a _{mn}$ and 1-forms
$\x ^m_\a , \ti \x_m ^\a$.
The geometry in each patch is
given  in term of these by
(\ref{gis}),(\ref{hiso}) (with the definitions (\ref{biso}),(\ref{ftiiss}),(\ref{vbis}))
and these give a globally defined metric and 3-form as a result of (\ref{sdflkhsd}),(\ref{btrnnn}). For example, 
$B=\ha   B_{mn}
   \x ^m \we  \x ^n 
$ is a globally-defined 2-form as $B^\a = B^\b$.
   
In each patch,    $U_\a \simeq \bar U_\a \times T^{2d}$, the space of orbits under the action of $\hat k_m ^\a$ 
can be thought of  as    $\ti U_\a \simeq \bar U_\a \times T^{d}$ with fibre coordinates  $\ti X_m$. With the transition functions (\ref{xitrans}), these patch together to give the dual space $\ti M$.
This is the dual affine torus bundle with the $\ti L$ in the transition functions (\ref{transx}) for $M$ replaced with 
$L$ in the transition functions (\ref{xitrans}) for $ \ti M$. 
T-duality in each patch   acts through
(\ref{x-x}),(\ref{e-e})
and lead to  a dual metric $\ti g^\a$ and 3-form $\ti H^\a$ in $U_\a$ given by
(\ref{tgd}),(\ref{thd}), and these patch together to give a globally defined metric and 3-form on $\ti M$.

\section{Torus Fibrations with B-Shifts}

\subsection{B-Shifts with Killing Vectors}

Returning to the set-up of section 4,
suppose $(M,g,H)$ has $d$ globally defined Killing vector fields
$k_m$, with $\i_m\i_n \i_pH=0$ but suppose that
$\i_m\i_nH$ is not necessarily exact.
Then in each patch $U_\a$ there is a $B_{mn} ^\a$
with
\begin{equation}
\label{}
\i_m\i_nH=dB_{mn} ^\a
\end{equation}
and as $\i_m\i_nH$ is globally-defined, in overlaps 
$U_\a\cap U_\b$, $B_{mn} ^\a - B_{mn}^\b$ is closed, so that \begin{equation}
B_{mn} ^\a = B_{mn}^\b + c _{mn}^{\a\b}
\end{equation}
for some constants $c _{mn}^{\a\b}
$.
Then the transition functions for $\Theta $ are changed from (\ref{thtrans}) to
\begin{equation}
\label{thtransb}
\Theta _{mn}^\a -\Theta _{mn}^\b=c _{mn}^{\a\b} - \i _m d\l ^{\a\b}_n
\end{equation}
As a result, the vector fields $\hat k$ defined by
(\ref{khatis})
are not globally defined, 
\begin{equation}
\label{kmix}
\hat k_m^\a =\hat k_m^\b + c _{mn}^{\a\b} \ti k ^n_\b
\end{equation}
The condition that $\hat k_m^\a, \ti k ^n_\a$   have compact orbits in each patch, so that $\hat M$ is a $T^{2d}$ bundle, imposes a quantization condition on 
  the
constants $c _{mn}^{\a\b}
$.
If $X^m \sim X^m + 2\p R$, $\ti X_m \sim \ti X_m + 2\p\ti  R$
for some $R,\ti R$ (with $\ti R= (2\p k R)^{-1}$ if the conditions of section 5 are imposed), 
 then
the quantization condition on the $c$ is that   $(R/\ti R)c _{mn}^{\a\b}$ are integers.

In section 7, transition functions on $M$ that mix the $ k$ among themselves were considered, so that $M$ is a torus bundle which is not principle,  
and (\ref{kmix}) gives a generalisation in which  transition functions on $\hat M$   mix the $\hat  k$ with the $\ti k$,  so that $\hat M$ is an affine  $T^{2d}$ bundle which is not principle.
Then although the vector fields $k_m$ are globally defined on $M$, the $\hat k_m$ are not globally defined on $\hat M$.
The 1-forms $\x$ have trivial transition functions $\x^\a = \x ^\b$, but
\begin{equation}
\label{}
\ti \x ^\a_m= \ti \x ^\b_m +c _{mn}^{\a\b}\x^n
\end{equation}
The transition functions for $E=G+B$ are then
\begin{equation}
\label{eshift}
E^\a = E^\b + c^{\a\b}
\end{equation}

The T-duality transformation (\ref{x-x}),(\ref{e-e}) can now be applied in any given  patch to give a dual geometry 
with moduli $\ti E^{mn}$  given by $\ti E^\a = (E^\a)^{-1}$ in $U_\a$.
If this is done in each patch, then
  the transition functions (\ref{eshift}) give the transition functions
\begin{equation}
\label{tiesh}
\ti E^\a = \ti E^\b   (1+ c^{\a\b} \ti E ^\b)^{-1}
\end{equation}
for $\ti E^\a = (E^\a)^{-1}$.
As a result, the geometries on each patch $(\ti U_\a, \ti g^\a, \ti H^\a)$ do not fit together to give 
a geometry on $\ti M$, as the transition functions for $\ti g^\a, \ti H^\a$ following from (\ref{tiesh})
do not give tensor fields on $\ti M$.
The transition functions
for $E$ (\ref{eshift}) are through an $O(d,d;\Z)$ transformation
(\ref{tetrans})
with 
 \begin{equation}\label{oddmb}
h^{\a\b}=\left(
\begin{array}{cc}
\1 & c^{\a\b}\\
0 & \1 \end{array}\right),
 \end{equation}
 while those for 
$\ti E$ (\ref{tiesh})
are an $O(d,d;\Z)$ transformation 
with 
\begin{equation}\label{oddmb}
\ti h^{\a\b}=\left(
\begin{array}{cc}
\1 & 0\\
c ^{\a\b} & \1 \end{array}\right),
 \end{equation}
 This is of the form $\ti h^{\a\b}=h_T h^{\a\b}h_T^{-1}$ where $h_T$ is the T-duality transformation (\ref{gT}),
 as expected from \cite{Hull:2004in}.
 Then $\hat M$ is a $T^{2d}$ bundle over $N$ which will in general have
   $O(d,d;\Z)$ monodromy
 of the form
 \begin{equation}\label{mond}
M(\g)=\left(
\begin{array}{cc}
\1 & N(\g)\\
0 & \1 \end{array}\right)
 \end{equation}
 round   1-cycles $\g$ in $\hat M$ for some integers $N(\g)$.
The transition functions are T-dualities, giving a T-fold \cite{Hull:2004in}.
Although the resulting background is not a conventional geometry on $\ti M$, it does give a good non-geometric background for string theory \cite{Hull:2004in}, as the transition functions are   a symmetry of string theory.

In this case, there are global issues in understanding the  T-duality   from the  point of view of the gauged sigma-model. 
In any given patch, the T-duality can be achieved   through gauging the 
isometries generated by $\hat k_m^\a$, giving a gauged lagrangian $L^\a$. However, these cannot be patched together to form a global gauged lagrangian as (\ref{kmix})
implies that the transition functions mix the isometries being gauged with those that are not.
Then the $T^d$ 
generated by the 
$\hat k$ do not  patch together to give a $T^d$ bundle over 
 $N$, and this leads to
the fact that the dual metric $\ti g$ and 3-form $\ti H$ are not globally-defined.
One might  instead attempt to gauge the   
isometries generated by $K_m^\a= \hat k_m^\a$ in $U_\a$ and the isometries
generated by $K^\b_m= \hat k_m^\b + c _{mn}^{\a\b} \ti k ^n_\b$ in $U_\b$, and in this way try to define 
globally defined vector fields $K_m^\a$ that can be gauged. 
However, there is a topological obstruction to doing this if
$\hat M$ has non-trivial $O(d,d;\Z)$ monodromy, i.e. if there is at least one 1-cycle $\g$ with $N(\g)\ne 0$.
If all monodromies are trivial, then one can construct a globally-defined 
$B_{mn}$ by taking $B_{mn}=
B_{mn} ^\a $ in $U_\a$, $B_{mn}= B_{mn}^\b + c _{mn}^{\a\b}$ in $U_\b$ etc and so
recover the set-up of section 4 with globally-defined $B_{mn}$.

\subsection{B-shifts and Torus Fibrations}

Consider now the situation of section 7 where $(M,g,H)$ is a torus fibration with local
Killing vector fields in each patch with transition functions (\ref{kpatch}), and supose $\i_m \i _n \i _p H=0$.
Then 
from (\ref{iihh}))
\begin{equation}
\label{}
d(B^\a_{mn} -L_m{}^p L_n {}^q B^\b_{pq} )=0
\end{equation}
so that 
\begin{equation}
\label{}
B^\a_{mn} -L_m{}^p L_n {}^q B^\b_{pq}= c_{mn}^{\a\b}
\end{equation}
for some constants 
$c_{mn}^{\a\b}$.
The transition functions for the vector fields
$\hat k$ are now
\begin{equation}
\label{kmix}
\hat k_m^\a =(  L_{\a\b})_m{}^n\hat k_n^\b + c _{mn}^{\a\b} \ti k ^n_\b
\end{equation}
and  the constants  $c _{mn}^{\a\b}$ satisfy the same quantization condition as in the last section, so that the orbits of $\hat k, \ti k$ are compact on each patch.
The transition functions for the 1-forms are
\begin{eqnarray}
\x ^m_\a & = & (\ti L _{\a\b}) ^m {}_n\x^n _\b \nonumber \\
\ti \x _m ^\a &= &( L _{\a\b})_m{}^n\ti \x _n ^\b  +c _{mn}^{\a\b}\x^n
\end{eqnarray}
The transition functions are then through the $O(d,d;\Z)$ transformations
\begin{equation}\label{oddmba}
h^{\a\b}=\left(
\begin{array}{cc}
\ti L _{\a\b}& c^{\a\b}\\
0 & L _{\a\b}\end{array}\right),
 \end{equation}
 In each patch one $U_\a$ one  can  T-dualise
using the formulae of section 6.
This again gives a T-fold, with transition functions
$\ti h^{\a\b}=h_T h^{\a\b}h_T^{-1}$ with $h^{\a \b}$ given by (\ref{oddmba}).

\section{T-Folds and T-Duality}\label{T-Folds and T-Duality}

The backgrounds considered here and in \cite{Hull:2004in} 
are constructed from local patches that are each conventional geometric string backgrounds.
For torus fibrations, these patches are
 of the form
$U_\a \simeq \bar U_\a \times T^d$ where $\bar U_\a$ are patches on the base $N$. In each such patch, the background has a conventional geometry $(U_\a, g^\a, H^\a)$
and  $U_\a$ is assumed to have  $d$ vertical  Killing vector fields $k_m$ tangent to the torus fibres.
The geometry $(U_\a, g^\a, H^\a)$ is determined by a geometry
$(\bar U_\a, \bar g^\a, \bar H^\a)$
on the base patch $\bar U_\a$ with metric $\bar g^\a$ and 3-form $\bar H^\a$, together with $T^d$ moduli $E_{mn}^\a=G_{mn}^\a+B_{mn}^\a$
and the $U(1)^{2d}$ connections
$A^m_\a, \ti A_m^\a$. The $A^m$ are the $U(1)^d$ connections associated with the $T^d$ fibration.

It was seen in section 4 that it  is natural to use this data to construct a $T^{2d}$ fibration by introducing $d$ extra toroidal dimensions to construct a patch
$\hat U_\a \simeq \bar U_\a \times T^{2d}$ with $U(1)^{2d}$ connection 1-forms
$\ca _\a =(A^m_\a, \ti A_m^\a)$.
Then there are $2d$ 1-forms $\x ^m , \ti \x_m$ on $\hat U_\a$ whose horizontal projections are
$A^m_\a, \ti A_m^\a$, and
there are $2d$ Killing vector fields
$\hat k_m , \ti k^m$ tangent to the fibres.

If $\i_m\i_n \i_pH=0$, there is a natural action of
$O(d,d)$ on  the geometry, with $E$ transforming as  (\ref{tetrans}), $\ca=(A,\ti A)$ transforming as    (\ref{atranss}),  $\x ^m , \ti \x_m$
 transforming as   (\ref{xicol}),(\ref{xxitrans}) and $\bar g, \bar H$ invariant.
 The subgroup $O(d,d;\Z)$ is a symmetry of string theory, as two backgrounds related by
 $O(d,d;\Z)$ define the same quantum theory.

 The string background  $M$  is constructed by patching the $U_\a$ together.
 In overlaps $U_\a\cap U_\b$, the patching conditions relating $(E^\a, \ca ^\a)$ to $(E^\b, \ca ^\b)$
 are given by a $U(1)^{2d}$ gauge transformation together with an $O(d,d;\Z)$ transformation $h^{\a\b}$.
  The background is geometric if the metrics  $g^\a $ and 3-forms $H^\a$ patch together to give 
 a metric tensor and 3-form on $M$. This requires that
 all the $h^{\a\b}$ can be taken to be of the form
 (\ref{oddmba}), so that the monodromies are all in the geometric subgroup 
 $\G(\Z)$ of matrices of the form (\ref{gG}).
 The $k^\a$
will be globally-defined vector fields provided the transition functions are all of the form
 (\ref{oddmb}), so that the monodromies are in the subgroup of matrices of the form (\ref{gB}).
 For general  $\G(\Z)$ monodromies, $M$ is a $T^d$ bundle over $N$.

For $O(d,d;\Z)$ monodromies that are not in  $\G(\Z)$, $M$ is a T-fold. This can be viewed as 
a manifold $M$ on which the $g^\a $ and   $H^\a$ do not  patch together to give 
tensor fields on $M$.
Such T-folds are non-geometric backgrounds, but nonetheless can provide good string backgrounds \cite{Hull:2004in}.
The transition functions in 
$O(d,d;\Z) \ltimes U(1)^{2d}$ can be used to patch the $\hat U_\a$ together to form a $T^{2d}$ bundle
$\hat M$ over $N$ with connection $\ca$.
The $\hat k_m , \ti k^m$  will be globally-defined  vector fields on $\hat M$ only if the $O(d,d;\Z)$ monodromies are trivial.

The topology of the $T^{2d}$ bundle $\hat M$ over $N$ is characterised by the
$2d$ first Chern classes $[\cf]\in H^2(N,\Z)$ and the $O(d,d;\Z)$ monodromies $g(\g)$ round 1-cycles $\g $ in $N$.  An $O(d,d;\Z)$ T-duality transformation $h$ on these is $[\cf] \to h^{-1} [\cf]$, $g(\g )\to h g(\g ) h^{-1}$.

The orbits of the $\hat k_m$ define a space $U'_\a \simeq \bar U_\a \times 
T^d \subset \hat U_\a$, and these patch together to form a $T^d$ bundle over $N$ if the monodromies are all in
the $GL(d,\Z)$ subgroup. In that case, if $\i_m\i_n \i_pH=0$ there is a gauged sigma-model on $\hat M$
in which the action of the $\hat k_m$ is gauged,
 and this can be used to show that the action of the T-duality group $O(d,d;\Z)$ on the geometry
is a symmetry of the quantum theory, 
and it takes a geometric background with $GL(d,\Z)$  monodromies to
 a geometric background with $GL(d,\Z)$  monodromies.
This extends the proof of T-duality to the case of torus fibrations with $GL(d,\Z)$ monodromies, and this is the maximal case in which a complete proof can be given in the way discussed here using a globally-defined gauged sigma-model. The condition that  the monodromies are all in
  $GL(d,\Z)$  is equivalent to the condition that  $\i_m\i_n H$ is exact.

In the general case, one can construct a gauged sigma-model in any patch $\hat U_\a$ in which the symmetry generated by the $\hat k$ is gauged provided  $\i_m\i_n \i_pH=0$, and this can be used to construct a dual geometry
$(\ti U_\a , \ti g^\a , \ti H^\a)$ on the space of orbits $\ti U_\a \simeq \bar U_\a \times T^d$.
For physical effects localised within  $\hat U_\a$, the sigma model on the original geometry
$( U_\a ,  g^\a ,  H^\a)$ and the dual geometry $(\ti U_\a , \ti g^\a , \ti H^\a)$
 give equivalent quantum theories, so one  can in principle use either.
This dualisation can then be done in all patches.
If the original background had transition functions $h^{\a\b} \in O(d,d;\Z)$, the dual one
  has transition functions $\ti h^{\a\b} \in O(d,d;\Z)$ given by
$\ti h^{\a\b}=h_T h^{\a\b}h_T^{-1}$.
If the original space was a geometric background with monodromies in $\G(\Z)$ with non-trivial $B$-shifts, so that the monodromies are not all in $GL(d,\Z)$, the dual background
is a non-geometric T-fold. 
A discussion of T-duality in this general case can be given using the doubled formalism of \cite{Hull:2004in}; this will be discussed in a separate publication.

The most general case requires the relaxation of the constraint $\i_m\i_n \i_pH=0$, so that $\i_m\i_n \i_pH $
gives constants in each patch, and the algebra of the Killing vectors $\hat k, \ti k$ becomes non-abelian.
The results of \cite{Dabholkar:2005ve} suggest that T-duality should generalise to this case, but the non-abelian structure leads to issues similar to those that arise in non-abelian duality \cite{delaOssa:1992vc},\cite{Alvarez:1994zr},\cite{Giveon:1993ai}, so that the approach used here appears difficult to implement in that case.

 \section*{Acknowledgments}

I would like to thank Dan Waldram and Martin Ro\v cek for useful  
discussions.

\end{document}